\documentclass[letterpaper,twocolumn,10pt]{article}
\usepackage{usenix}

\usepackage{tikz}
\usepackage{amsmath}

\usepackage{graphicx}
\usepackage{multirow}
\usepackage{makecell}
\usepackage{booktabs}
\usepackage{enumitem}
\usepackage{pifont}
\usepackage{mdframed}
\usepackage{colortbl}
\usepackage{array}
\usepackage{adjustbox}
\usepackage{tabularx}
\usepackage{textcomp}
\usepackage{listings}
\usepackage[most]{tcolorbox}

\newif\ifshowrevisions
\showrevisionsfalse
\newcommand{\rev}[1]{\ifshowrevisions\textcolor{red}{#1}\else#1\fi}

\newtcolorbox{insightbox}[1][]{
  enhanced,
  breakable,
  colback=gray!10,
  colframe=gray!60,
  boxrule=0pt,
  leftrule=2pt,
  sharp corners,
  boxsep=0pt,
  top=5pt,
  bottom=5pt,
  left=5pt,
  right=5pt,
  fontupper=\normalsize,
  before skip=6pt,
  after skip=6pt,
  #1
}

\definecolor{verylightgray}{rgb}{.97,.97,.97}
\lstdefinelanguage{Solidity}{
  keywords=[1]{def, anonymous, assembly, assert, balance, break, call, callcode, case, catch, class, constant, continue, constructor, contract, debugger, default, delegatecall, delete, do, else, emit, event, experimental, export, external, false, finally, for, function, gas, if, implements, import, in, indexed, instanceof, interface, internal, is, length, library, log0, log1, log2, log3, log4, memory, modifier, new, payable, pragma, private, protected, public, pure, push, require, return, returns, revert, selfdestruct, solidity, storage, struct, suicide, super, switch, then, this, throw, transfer, true, try, typeof, using, value, view, while, with, addmod, ecrecover, keccak256, mulmod, ripemd160, sha256, sha3}, 
  keywordstyle=[1]\color{blue}\bfseries,
  keywords=[2]{address, bool, byte, bytes, bytes1, bytes2, bytes3, bytes4, bytes5, bytes6, bytes7, bytes8, bytes9, bytes10, bytes11, bytes12, bytes13, bytes14, bytes15, bytes16, bytes17, bytes18, bytes19, bytes20, bytes21, bytes22, bytes23, bytes24, bytes25, bytes26, bytes27, bytes28, bytes29, bytes30, bytes31, bytes32, enum, int, int8, int16, int24, int32, int40, int48, int56, int64, int72, int80, int88, int96, int104, int112, int120, int128, int136, int144, int152, int160, int168, int176, int184, int192, int200, int208, int216, int224, int232, int240, int248, int256, mapping, string, uint, uint8, uint16, uint24, uint32, uint40, uint48, uint56, uint64, uint72, uint80, uint88, uint96, uint104, uint112, uint120, uint128, uint136, uint144, uint152, uint160, uint168, uint176, uint184, uint192, uint200, uint208, uint216, uint224, uint232, uint240, uint248, uint256, var, void, ether, finney, szabo, wei, days, hours, minutes, seconds, weeks, years},  
  keywordstyle=[2]\color{teal}\bfseries,
  keywords=[3]{block, blockhash, coinbase, difficulty, gaslimit, number, timestamp, msg, data, gas, sender, sig, value, now, tx, gasprice, origin},  
  keywordstyle=[3]\color{violet}\bfseries,
  keywords=[4]{[1]},
  keywordstyle=[4]\color{blue}\bfseries,
  identifierstyle=\color{black},
  sensitive=false,
  comment=[l]{//},
  morecomment=[s]{/*}{*/},
  commentstyle=\color{gray}\ttfamily,
  stringstyle=\color{red}\ttfamily,
  morestring=[b]',
  morestring=[b]"
}
\begin{document}

\date{}


\title{\Large \bf Bridging the Opacity: Evidence-Backed Cross-Chain Transaction Correspondence Reconstruction Across Heterogeneous Blockchains}

\author{
Dan Lin$^{1}$ \quad
Huan Xiao$^{1}$ \quad
Ziwei Li$^{1}$ \quad
Xiapu Luo$^{2}$ \\
Jiachi Chen$^{3}$ \quad
Jiajing Wu$^{1}$ \quad
Zibin Zheng$^{1}$ \\[4pt]
$^{1}$Sun Yat-sen University \quad
$^{2}$The Hong Kong Polytechnic University \quad
$^{3}$Zhejiang University
}

\maketitle

\begin{abstract}
\rev{Cross-chain bridges enable interoperability, but they also break the transaction trails needed to trace illicit funds. Third-party investigators typically cannot access the source-to-destination mappings maintained by bridge backends, and our survey of 131 bridges finds that only 16.79\% provide complete public tracking. Existing approaches depend on official APIs, EVM-specific assumptions, or fragile temporal heuristics, limiting their ability to trace transfers across heterogeneous ledgers. We present \textsc{XSplicer}, an evidence-driven system for reconstructing \textit{cross-chain transaction correspondence (xTCR)} without privileged access to bridge backends. \textsc{XSplicer} derives unified semantic specifications from public protocol documentation and transaction examples, translates them into lightweight parsers and verifiers, and links source and destination transactions by prioritizing hard evidence and using soft clues only when necessary.}
\rev{We evaluate \textsc{XSplicer} on seven bridge protocols spanning EVM, Bitcoin, and Solana. \textsc{XSplicer} achieves 92.5\% global recovery rate and up to 98.61\% on individual protocols. Under adversarial noise, its hard-evidence verifier retains the correct match in 100\% of tested cases, while soft-clue matching degrades as ambiguity increases. In two real-world case studies, \textsc{XSplicer} recovers more than 1,900 historical transaction pairs after Multichain ceased operations and identifies 754 illicit cross-chain transfers worth \$105.6 million in the Bybit laundering incident. These results show that public protocol invariants can support practical cross-chain forensics without privileged bridge mappings.}


\end{abstract}

\section{Introduction}
\label{sec:intro}

\rev{As blockchain networks evolve from isolated value islands into interconnected heterogeneous networks~\cite{belchior2021survey,zamyatin2021sok,li2025interoperability}, cross-chain bridges have become key infrastructure for asset interoperability~\cite{augusto2024sok,han2023survey}. As of our access date, DeFiLlama~\cite{defillama_bridges} reports roughly \$380M in 24-hour bridge volume and \$15B over one month, placing bridge-mediated transfers at a hundreds-of-millions-of-dollars daily scale in decentralized finance~\cite{gramlich2023multivocal,kitzler2023defi}.}

However, while this interoperability provides significant convenience, it disrupts the tracking continuity inherent in single-chain environments, severely compromising the observability of transactions. Attackers leverage ``chain-hopping''~\cite{yousaf2019tracing,elliptic2025} techniques to easily sever tracking leads on isolated ledgers, dispersing illicit funds across heterogeneous networks. According to the Elliptic 2025 report~\cite{elliptic2025}, over \$21.8 billion in high-risk on-chain assets have been transferred via cross-chain mechanisms. For third-party analysts (e.g., security firms and regulators) constrained by publicly available data, cross-chain bridges are increasingly evolving into observability blind spots that obstruct tracking links.

\begin{figure}
    \centering
    \includegraphics[width=1\linewidth]{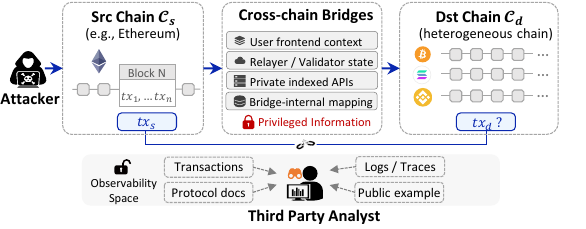}
    \caption{Limited observability in cross-chain transaction correspondence reconstruction. An attacker initiates a source-chain transaction $tx_s$  that is bridged to a destination chain, resulting in an unknown destination transaction $tx_d$.}
    \label{fig:Introduction}
\end{figure}

To quantify the extent of this observability deficit, we conduct an empirical analysis of 131 mainstream cross-chain protocols within the DeFi ecosystem. The results indicate that only 16.79\% of the investigated bridge protocols provide precise destination-transaction $tx_d$ retrieval capabilities based on source transaction hashes $tx_s$ (detailed in Section~\ref{sec:background:observability}). This implies that for the vast majority of cross-chain activities, third-party analysts are deprived of explicit on-chain mapping evidence, as shown in Fig.~\ref{fig:Introduction}. Consequently, when official bridge interfaces are unavailable, oracle services are disrupted, or a protocol completely ceases operations (e.g., the Multichain shutdown~\cite{multichain_news} in 2023), investigators are forced to operate without any official auxiliary indexing. In such scenarios, they must rely solely on discrete state transition logs and transaction payloads scattered across heterogeneous chains. However, these isolated records inherently lack direct semantic continuity at the architectural level, imposing prohibitive analytical barriers to cross-chain forensics.

\rev{Prior work~\cite{yousaf2019tracing,lin2025connector,lin2025abctracer} uses heuristics to bound the search space when bridge-internal mappings are invisible, but two core challenges remain.} First, the heterogeneity of cross-chain protocols and execution environments exacerbates semantic discontinuity at the architectural level. Existing cross-chain association methods predominantly focus on relatively homogeneous scenarios (e.g., Ethereum $\to$ Polygon), relying on parsing logic built upon the assumption of similar address formats and transaction structures. However, within contemporary heterogeneous cross-chain architectures, destination clues are frequently fragmented and re-encoded within underlying payloads. For instance, in a deBridge (Ethereum $\to$ Solana) transaction, a 32-byte hexadecimal destination address on the Ethereum side must undergo protocol-specific \texttt{Base58} parsing to map to a native Solana public key (detailed in Section~\ref{sec:overview:Motivation}). \rev{This heterogeneous re-encoding requires protocol-specific decoding and normalization before the relevant fields can be compared across ledgers.} Second, existing approaches over-rely on probabilistic features such as time and amount similarities, lacking the evidentiary rigor required for forensics. In high-throughput or adversarial scenarios, these methods are highly susceptible to generating massive candidate pools and inducing false associations.

\begin{figure*}[t]
    \centering
    \includegraphics[width=0.95\linewidth]{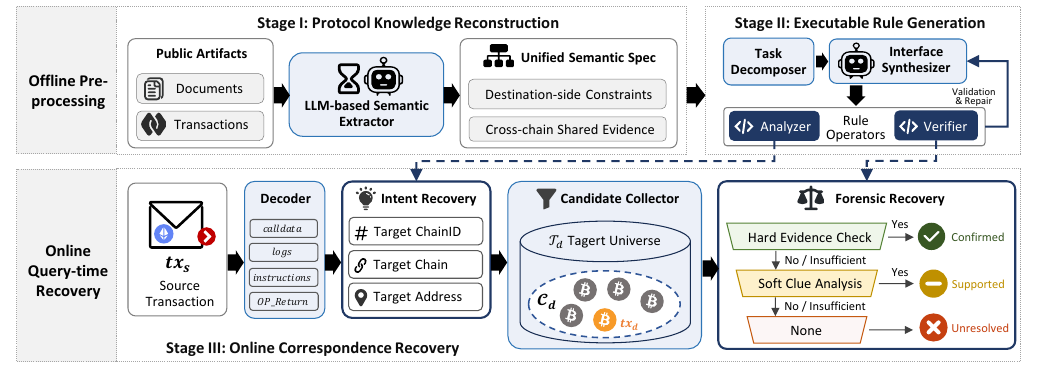}
    \caption{\rev{Overview of the \textsc{XSplicer} architecture. Offline, Stage I derives a unified semantic specification of destination-side constraints and cross-chain shared evidence from public documents and transaction examples. Stage II compiles the specification into two lightweight executable rule operators: the \texttt{Analyzer} and the \texttt{Verifier}. Online, Stage III uses the \texttt{Analyzer} to construct a bounded candidate set and applies hard-evidence-first verification to produce a forensic decision state.}}
    \label{fig:architecture}
\end{figure*}

To address the challenge of cross-chain transaction correspondence reconstruction under non-privileged conditions, we introduce \textsc{XSplicer}, as illustrated in Fig.~\ref{fig:architecture}. \rev{\textsc{XSplicer} implements protocol-evidence compilation as a three-stage workflow. Stage I uses Large Language Models (LLMs)~\cite{brown2020language,wei2022chain,lewis2020retrieval,yao2023react} to synthesize public protocol documents and execution traces into a unified semantic specification of destination-side constraints and cross-chain shared evidence. Stage II compiles this specification into an executable \texttt{Analyzer} and \texttt{Verifier}. Stage III uses these operators to bound destination candidates and assign an evidence-linked forensic decision. This separation of candidate construction and verification prioritizes protocol-level hard evidence and uses soft clues only as fallback support when hard evidence is unavailable.}

\rev{We evaluate \textsc{XSplicer} on seven mainstream bridges across Ethereum, Bitcoin, and Solana, achieving accuracy ranging from 87.92\% to 98.61\%. The results show that public protocol evidence supports correspondence reconstruction across heterogeneous execution models, even under constrained third-party observability. Comparisons with existing tools and official interfaces expose systemic deficiencies in the interfaces' historical coverage and handling of non-standard protocol states (e.g., cross-chain attacks). Component analyses and robustness experiments further show that protocol-level invariants retain discriminative power under candidate ambiguity and adversarial noise, whereas time- and amount-based soft clues degrade under interference. Case studies further demonstrate practical forensic utility. \textsc{XSplicer} reconstructs historical paths after the Multichain shutdown and traces THORChain-mediated flows associated with the Bybit Hack.}
Our primary contributions are summarized as follows.

\begin{itemize}[leftmargin=*]\itemsep0.6pt
    \item \rev{\textbf{Problem formulation.} We formulate xTCR for third-party investigators as candidate search guided by destination-side constraints, followed by verification using shared protocol evidence.}
    \item \rev{\textbf{System design.} We design and implement \textsc{XSplicer}, which compiles public protocol artifacts into semantic specifications and executable analyzers and verifiers for heterogeneous ledgers.}
    \item \rev{\textbf{Evaluation.} We evaluate \textsc{XSplicer} on 54,578 transaction pairs across seven mainstream bridges spanning EVM, Bitcoin, and Solana, including adversarial noise experiments and real-world case studies.}
    \item \rev{\textbf{Forensic findings.} We show that official bridge interfaces can miss valid correspondences and that protocol-level hard evidence resolves ambiguities left by time- and amount-based matching.}
\end{itemize}

\section{Background and Empirical Study}
\label{sec:background}

\subsection{Cross-Chain Bridge Workflow}
\rev{Bridge workflows generally follow three stages~\cite{augusto2024sok,han2023survey,debridge_lifecycle}. \textcircled{\small 1} During initiation, the source protocol \textit{locks}, \textit{burns}, or \textit{escrows} assets and emits a cross-chain event or message containing user information. \textcircled{\small 2} During relaying, validators or relayers propagate the signed proof to the destination environment. \textcircled{\small 3} During settlement, the destination protocol validates the proof and \textit{mints} or \textit{releases} assets. We use the following transaction definitions:}

\begin{itemize}[leftmargin=*]\itemsep0.6pt
    \item \textbf{Source transaction ($tx_{s}$)}: The transaction on the source chain that initiates the bridge request and asset transfer.
    \item \textbf{Destination transaction ($tx_{d}$)}: The transaction that concludes the process via asset delivery on the destination chain. Any auxiliary transactions, including verification proofs, intermediate hops, or \textit{refund/return} executions, are categorized as related transactions ($tx_{rel}$).
\end{itemize}
In a third-party setting, these transactions represent the only observable anchors for reconstructing fund flows across disjoint ledgers.

\subsection{Blockchain Heterogeneity}
\rev{Heterogeneous execution models encode and store cross-chain evidence differently. EVM chains distribute it across \textit{calldata}, \textit{event logs}, and internal \textit{traces}; Bitcoin embeds it in locking scripts, change outputs, or \texttt{OP\_RETURN} payloads; and Solana stores it in multi-instruction transactions, account metadata, and program logs. The same evidence therefore appears in different formats and locations across chains, making simple heuristic matching insufficient.}

\subsection{Empirical Study: The Observability Crisis}
\label{sec:background:observability}

To quantify the forensic visibility of existing cross-chain bridges, we conduct an empirical investigation of 131 mainstream cross-chain protocols and bridge-integrated DeFi services sourced from the cross-chain aggregator Chainspot~\cite{chainspot_list}. We define \textit{\textbf{traceability}} as the consistent availability of a mapping from a source transaction hash ($tx_{s}$) to its corresponding destination transaction hash ($tx_{d}$) via public official interfaces.

Our findings reveal a pervasive observability crisis: only 16.79\% of the surveyed protocols provide such verifiable mapping capabilities. Due to space constraints, the complete list of 131 bridges is presented in TABLE~\ref{tab:bridge-survey-two-panel} of Appendix~\ref{app:data:bridge-survey}. 

Furthermore, most public APIs are inherently designed for developer-oriented integration, pricing, and real-time status tracking, rather than for forensic-grade historical record reconstruction. For instance, the LayerZero Value Transfer API requires an application for a production environment API key to gain access~\cite{layerzero_value_transfer_api}. The documentation for the Across protocol explicitly positions its API as a tool for generating executable \texttt{calldata} and quotes expressly advising developers against caching response data~\cite{across_api}. Explorer interfaces such as Wormholescan primarily offer superficial metadata browsing functionalities instead of comprehensive and independently verifiable forensic indices intended for external investigators~\cite{wormholescan_api}. Following the cessation of operations by Multichain in 2023, all its APIs and the official explorer became inaccessible for queries. Similarly, the deprecation of the legacy API for Stargate underscores the volatility of such interfaces~\cite{stargate_api_deprecated}.

Therefore, official APIs cannot be treated as reliable correspondence oracles. In a third-party analysis setting, cross-chain correspondence must be performed without bridge-internal state and must rely on public on-chain artifacts and protocol documentation. Since temporal and monetary proximity provide only soft cues and lack the evidentiary strength required for forensic attribution, \textsc{XSplicer} moves beyond similarity-based heuristics and reconstructs correspondence from verifiable hard evidence.

\section{Problem Overview}

Motivated by the privileged invisibility and representational heterogeneity observed in Section~\ref{sec:background:observability}, this part defines the problem model for third-party analysis. We specify the capabilities of the analyst and the adversary, and \textbf{extend prior cross-chain association analysis to \textit{cross-chain transaction correspondence reconstruction (xTCR)}}, where the goal is to recover evidence-backed source--destination correspondence under privilege barriers.

\subsection{Third-Party Analyst Model}
\rev{We define the analyst as a non-privileged external entity, representing security auditors or regulatory bodies conducting post-mortem forensics~\cite{yousaf2019tracing,lin2025connector,lin2025abctracer}. The analyst can access public evidence, including on-chain artifacts, public protocol specifications, and publicly available cross-chain examples. We assume no access to bridge-internal privileged information, such as private backend mappings or proprietary relayer states, and no continuously available official correspondence service. This model targets non-cooperative and post-mortem investigations; when official mappings are available, \textsc{XSplicer} provides an independent cross-check rather than replacing cooperative access.}

\subsection{Adversary Model}
\rev{We consider attackers who use ``chain-hopping''~\cite{yousaf2019tracing,elliptic2025} to disrupt single-chain tracing by moving assets to a heterogeneous chain, aiming to obscure the public link between the source transaction ($tx_s$) and its destination transaction ($tx_d$). Attackers may also vary transfer amounts to hinder recovery. We assume an honest-but-opaque bridge that may maintain a deterministic internal mapping inaccessible to third-party analysts. Thus, xTCR addresses observability loss rather than a missing mapping or bridge compromise.}

\subsection{Problem Formulation}

We model cross-chain transaction linkage as the reconstruction of latent transaction correspondence under a privilege barrier. The input is $\langle\mathcal{C}_s, tx_s, \mathcal{A}, \mathcal{T}_s\rangle$, where $tx_s \in \mathcal{T}_s$ is a publicly observable cross-chain initiation transaction on the source chain $\mathcal{C}_s$; $\mathcal{A}$ denotes the public artifact space of the public protocol, including unstructured documentation and a limited set of pairs of public cross-chain examples; and $\mathcal{T}_d$ denotes the complete set of public transactions on the destination chain $\mathcal{C}_d$. Because $\mathcal{T}_s$ and $\mathcal{T}_d$ differ in their underlying execution models, such as Account-based and UTXO-based models~\cite{ethereum_transactions,bitcoin_transactions}, the destination-side execution transaction $tx_d$ of $tx_s$ is latent. Before reconstructing the protocol logic, an analyst cannot directly identify the
 deterministic attributes of $tx_d$, such as its hash or address, from the full set $\mathcal{T}_d$. We therefore construct an evidence reasoning model based on bridge protocol invariants to address this heterogeneity-induced observability gap.
The output is $y=\langle\mathcal{C}_d, (tx_s,tx_d), E\rangle$, where $tx_d \in \mathcal{T}_d$ is the reconstructed destination transaction, i.e., the transaction that completes the delivery of the user's asset, and $E$ is the evidence set supporting the correspondence. If no correspondence can be established, then $y=\emptyset$. Each evidence item $e_i \in E$ is a multidimensional tuple that explicitly states the logical link between $tx_s$ and $tx_d$. Section~\ref{sec:system:stage1} defines this structure as a specification in detail.

\subsection{Motivation: Why Prior Work Breaks Down}
\label{sec:overview:Motivation}
The limitations of existing cross-chain association methods in real-world analysis primarily stem from the fact that their underlying assumptions do not hold in our setting.

\begin{figure}[t]
    \centering
    \includegraphics[width=1\linewidth]{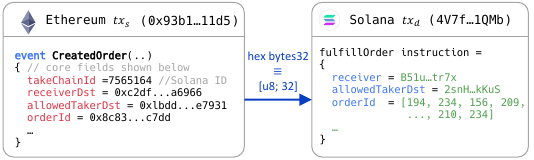}
    \caption{The representation gap in deBridge cross-chain transactions from Ethereum to Solana. Although semantically identical, a shared identifier is encoded as \texttt{hex bytes32} on Ethereum and as a \texttt{[u8;32]} byte array on Solana, making the semantic correspondence obfuscated.}
    \label{fig:motivation_debridge}
\end{figure}

First, in heterogeneous cross-chain scenarios, the task is no longer to simply select the most similar transaction from a candidate pool. Instead, the destination-side clues themselves elude existing parsing assumptions. As illustrated in Fig.~\ref{fig:motivation_debridge}, we examine a deBridge~\cite{debridge_dln_overview,debridge_dln_contracts,ethereum_transactions,solana_transactions} transaction from Ethereum \href{https://etherscan.io/tx/0x93b16cc2876e7aa8f30d1e294132a238139d90cb2256169c263ebf2bcbaa11d5}{\texttt{0x93b1...11d5}} to Solana \href{https://solscan.io/tx/4V7fau3r7qnxGEGapi2YTpskgMFMXuP9fVfHpHojuwxMdgVjpd9k4L9kjnrkMngygNfxkpdP62byvZC1qENV1QMb}{\texttt{4V7f...1QMb}}, in order to demonstrate this representation gap. While the \texttt{CreatedOrder} event on the source chain should logically correspond to the \texttt{fulfillOrder} instruction on the destination chain, this relationship lacks direct observability. On the Ethereum side, destination information is embedded within internal fields rather than explicitly exposed. The \texttt{takeChainId} parameter designates Solana as the target, but critical fields like \texttt{receiverDst} and \texttt{allowedTakerDst} are encoded as \texttt{bytes32} values that cannot be directly interpreted as Solana addresses. Similarly, the cross-chain identifier \texttt{orderId} manifests as a \texttt{hex bytes32} string on Ethereum, whereas it is represented as a \texttt{u8} array of size 32 within the Solana instruction. Although these elements exhibit semantic consistency, their on-chain representations are entirely heterogeneous, precluding direct alignment through string or structural matching. Consequently, recovering the correspondence fundamentally requires decoding the destination-side fields from the source event into valid Solana public keys and normalizing the \texttt{orderId} into a unified byte representation to ensure robust verification.

Second, even when cross-chain transfers occur in seemingly homogeneous environments, the temporal and monetary similarities utilized by existing work can at best help shrink the candidate space. However, attack samples from THORChain demonstrate that a single source chain behavior can simultaneously correspond to multiple seemingly valid destination-side candidates, depriving monetary proximity of its uniqueness capability. Furthermore, in directions like BTC$\to$ETH, long delays may occur between the source chain transaction and the destination chain settlement, significantly expanding the search window and introducing more similar-amount interference items (detailed analysis in Section~\ref{sec:comparison:bridge_interfaces}).

These two types of failures collectively indicate that, in a third-party analyst setting, the key to cross-chain reconstruction is no longer finding the most similar destination chain transaction for each source chain transaction. Instead, the key is to first restore the comparability of destination-side constraints and then verify protocol-level shared evidence within a local candidate space.

\subsection{Challenges and Design Rationale}

The failures outlined above introduce three structural constraints that fundamentally shape our methodology.

\noindent\textbf{Challenge I:} The opacity of bridge-internal mappings. Without privileged access, the $tx_s \to tx_d$ mapping is invisible to external analysts. In the absence of internal APIs that bound the search space, analysts face a severe observability gap.

\noindent\textbf{Design I:} Artifact-driven semantic reconstruction. \textsc{XSplicer} does not start by blindly scanning ledgers. It first extracts specifications, including destination intent and shared evidence, from public artifacts and examples. This provides a declarative basis for reconstruction without privileged information.
    
\noindent\textbf{Challenge II:} Representation heterogeneity. Different execution environments fragment and re-encode cross-chain clues in low-level payloads, breaking their direct comparability and making static, template-based parsing insufficient.

\noindent\textbf{Design II:} Executable specification synthesis. To bridge this structural gap, \textsc{XSplicer} dynamically compiles reverse-engineered protocol semantics into executable decoding operators and normalization rules, enabling uniform comparison across heterogeneous ledgers.
    
\noindent\textbf{Challenge III:} The inconclusiveness of soft clues. Public traces are inherently ambiguous. Temporal and monetary heuristics can conflate statistical correlation with factual, verifiable correspondence.

\noindent\textbf{Design III:} Evidence-stratified verification. The verification engine follows a hard-evidence-first principle. Soft clues are used only as probabilistic fallback when structural evidence is unavailable. This allows \textsc{XSplicer} to output evidence-backed forensic states rather than fragile similarity rankings.


\section{System Design}
\label{sec:system}

\begin{figure}[t]
    \centering
    \includegraphics[width=0.95\linewidth]{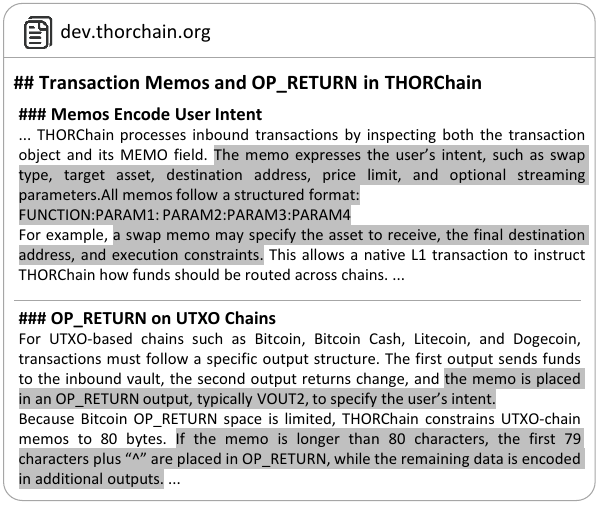}
    \caption[A public documentation snippet from THORChain Dev Docs.]{
    A public documentation snippet from THORChain Dev Docs~\cite{thorchain_dev_docs} serving as input for Stage I. The highlighted text describes how transaction memos and OP\_RETURN fields encode user intent, destination addresses, routing constraints, and execution parameters for cross-chain swaps.
    }
    \label{fig:thorchain_Docs}
\end{figure}

\rev{\textsc{XSplicer} provides a unified workflow for evidence-backed correspondence reconstruction across heterogeneous ledgers.} 
As illustrated in Fig.~\ref{fig:architecture}, \textsc{XSplicer} adopts a three-stage system architecture.

\textbf{Stage I: Protocol knowledge reconstruction.} \textsc{XSplicer} first reconstructs protocol-level semantics from public artifacts, including protocol documents and publicly available cross-chain transaction examples. The goal of this stage is not to directly identify the destination transaction, but to derive a unified semantic specification that captures two types of recoverable hard evidence: destination-side constraints and cross-chain shared evidence. Destination-side constraints (e.g., the destination chain and recipient) define where the corresponding destination-side execution may appear. Cross-chain shared evidence (e.g., message identifiers and source-transaction references) provides the basis for later verification. This intermediate specification provides a structured interface between heterogeneous public evidence and executable recovery logic.

\rev{\textbf{Stage II: Executable rule generation.} Given the unified semantic specification, \textsc{XSplicer} synthesizes two lightweight executable rule operators. The \texttt{Analyzer} extracts destination-side constraints from a decoded source transaction for online recovery, while the \texttt{Verifier} extracts, normalizes, and checks cross-chain shared evidence between a source transaction and a destination-side candidate. Rather than generating a general-purpose bridge analyzer, Stage II translates the recovered protocol semantics into bounded, auditable interfaces between high-level protocol descriptions and low-level chain-specific encodings.}

\textbf{Stage III: Online correspondence recovery.} Given a source chain transaction $tx_s$, \textsc{XSplicer} first decodes data, then the \texttt{Analyzer} recovers destination-side constraints and uses them to construct a bounded candidate set $\mathcal{K}_d$ from the destination-side transaction universe $\mathcal{T}_d$. After that, the \texttt{Verifier} applies a hard-evidence-first decision strategy over $\mathcal{K}_d$. If normalized cross-chain shared evidence uniquely supports a candidate, \textsc{XSplicer} outputs a confirmed correspondence. If hard evidence is unavailable but soft clues provide auxiliary support, the system outputs a supported result rather than a proof.

\subsection{Stage I: Protocol Knowledge Reconstruction}
\label{sec:system:stage1}

The objective of Stage I is to synthesize machine-actionable protocol rules from heterogeneous public artifacts, without privileged correspondence oracles. Our design relies on the observation that cross-chain correspondence appears in two complementary forms:
\begin{itemize}[leftmargin=*]\itemsep0.6pt
    \item \textit{Textual protocol artifacts} (e.g., official documentation and developer guides), which describe the declarative encoding of destination intent.
    \item \textit{On-chain execution artifacts} (e.g., public source--destination examples), which show how these semantics appear in native ledger structures, such as EVM logs and Solana instructions.
\end{itemize}
Stage I synthesizes these two modalities into a unified semantic specification (Semantic Spec).

\textbf{LLM-based semantic extractor.} To reduce noise and handle structural disparities, Stage I first applies targeted preprocessing to the input artifacts: 
\begin{itemize}[leftmargin=*]\itemsep0.6pt
    \item Documentation-side block filtering: \textsc{XSplicer} segments protocol documentation into discrete text blocks. We use an LLM as a semantic filter to retain blocks relevant to cross-chain payloads and shared evidence, while discarding irrelevant boilerplate, such as SDK installation guides.
    \item Example-side trace summarization: For verified examples, \textsc{XSplicer} uses offline parsing operators to convert raw low-level transaction traces into structured representations, which helps the LLM better interpret empirical evidence.
\end{itemize}

\rev{\textbf{Public-example grouping and selection.} To cover different protocol execution patterns, we group and select public examples according to transaction structures on the source ledger. EVM examples are grouped by \texttt{methodId} and interacting contract addresses, whereas Solana examples are grouped by \texttt{programId} and instruction byte length. For Bitcoin examples, we use random selection by default or coarse grouping by \texttt{OP\_RETURN} length. Section~\ref{sec:component:input} evaluates how public examples affect performance.}

Second, \textsc{XSplicer} extracts hard-evidence patterns from the preprocessed data to populate the protocol \textit{Semantic Spec}:
\begin{itemize}[leftmargin=*]\itemsep0.6pt
    \item \textit{Destination-side constraints:} These include destination chain IDs, address-encoding variants, and recipient boundaries, which delimit the downstream search space.
    \item \textit{Cross-chain shared evidence:} These are deterministic invariants preserved across ledgers, such as unique message IDs, nonces, or protocol-specific hash references.
\end{itemize}
Stage I then merges evidence extracted from documents and traces into a unified Semantic Spec. The merge procedure follows a conservative consensus rule: fields and logic are finalized only when corroborated by multiple independent evidence sources. Conflicting or weakly supported items are marked as pending. This cross-validated merging strategy reduces LLM-induced errors in protocol interpretation.

\begin{figure}[t]
    \centering
    \includegraphics[width=0.95\linewidth]{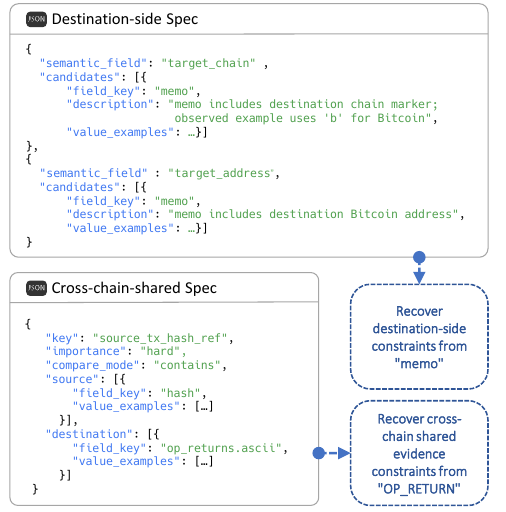}
    \caption{The unified semantic spec generated by Stage I for THORChain. The upper section instructs the extraction paths for destination-side constraints from the \texttt{memo} field, and the lower section specifies the cross-chain shared evidence from the \texttt{OP\_RETURN} payload. }
    \label{fig:Unified_Semantic_Spec}
\end{figure}

\textbf{Stage I example input and output.}
To illustrate how \textsc{XSplicer} operationalizes its evidence-driven reconstruction via the three-stage architecture, we walk through an end-to-end THORChain transfer (ETH$\to$BTC) as follows.
The system takes a public reference pair (ETH $tx_s$ \href{https://etherscan.io/tx/0x2b0557d7eb2c77c2524d911ccd8821a0550c5bb236ee0b9a93ce19a53d6665fc}{\texttt{0x2b05...65fc}} $\to$ BTC destination $tx_d$ \href{https://btcscan.org/tx/5BDCCCC3F8BE0727D586F2373AE47E04B239AAF94395550618DFFC1F5D0C5714}{\texttt{5BDC...5714}}), and THORChain's developer documentation~\cite{thorchain_memos,thorchain_sending_transactions} (as shown in Fig.~\ref{fig:thorchain_Docs}) as input. 
The unified semantic specification is generated as shown in Fig.~\ref{fig:Unified_Semantic_Spec}.
\textcircled{\small 1} \textit{Destination-side constraints}: The ETH transaction invokes \texttt{depositWithExpiry()}. The spec defines rules to extract the destination chain marker (e.g., \texttt{b} for Bitcoin) and the recipient address from the \texttt{memo}. 
\textcircled{\small 2} \textit{Cross-chain shared evidence}: The system identifies a rigid correlation invariant, the \texttt{OP\_RETURN} payload in the BTC destination transaction explicitly backfills the source transaction's hash (\texttt{source\_tx\_hash\_ref}). Because this reference is explicitly embedded by the destination protocol and uniquely points to the source execution, \textsc{XSplicer} categorizes it as protocol-level \textit{hard evidence}.

\subsection{Stage II: Executable Rule Generation}
\label{sec:system:stage2}

The semantic specifications produced in Stage I define what evidence should be recovered, but they do not directly parse the physical layout of raw on-chain payloads. Using LLMs for online transaction processing is unsuitable: the workload is large, latency-sensitive, and requires deterministic byte-level operations, such as address decoding, hash normalization, and offset-based field extraction. Stage II therefore converts the declarative Semantic Spec into executable rule operators that can be invoked by the online recovery engine.
As shown in Fig.~\ref{fig:architecture}, Stage II consists of two components: a \textit{task decomposer} and an \textit{interface synthesizer}. 

\textbf{Task decomposer.} 
The task decomposer first breaks the Unified Semantic Spec into two classes of executable tasks. The first class targets destination-side constraint recovery, including destination chain identifiers, destination addresses, recipient boundaries, and protocol-specific filters. The second class targets cross-chain shared evidence verification, including message IDs, nonces, hash references, and other protocol invariants that must be compared across ledgers.

\textbf{Interface synthesizer.} 
The interface synthesizer then materializes these tasks into two lightweight modules:
\begin{itemize}[leftmargin=*]\itemsep0.6pt
    \item \texttt{Analyzer:} The analyzer recovers destination-side constraints from decoded source-chain traces. It follows the semantic paths emitted by the Task Decomposer and applies the corresponding decoding and normalization rules, such as address-format conversion or field extraction from calldata, logs, memos, or instructions.
    \item \texttt{Verifier:} The verifier checks cross-chain shared evidence for a candidate transaction pair. It extracts and normalizes protocol-specific evidence from both sides and evaluates whether the pair satisfies the required invariants, such as hash concatenation, prefix padding, byte-order conversion, or sequence-number matching.
\end{itemize}

\rev{\textbf{Validation and repair.} After the Interface Synthesizer generates the Analyzer and Verifier, \textsc{XSplicer} automatically validates their executability, behavior on public cross-chain examples, and consistency with the Semantic Spec. If a check fails, \textsc{XSplicer} uses the validation report to guide the LLM in repairing the affected code and then revalidates the modules before they enter Stage III.}

Both modules expose standardized interfaces to Stage III. This separation keeps LLM-assisted reasoning offline and confines online recovery to deterministic operators. As a result, the runtime only applies fixed protocol-specific checks to each candidate pair, enabling scalable verification over large transaction sets.

\textbf{Stage II example output.} \textsc{XSplicer} statically compiles this Spec into executable code, as shown in Fig.~\ref{fig:Analyzer_Verifier}. 
\textcircled{\small 1} The \texttt{Analyzer} is programmed to parse the source-side \texttt{memo} to isolate the destination constraints. \textcircled{\small 2} The \texttt{Verifier} is synthesized to decode the \texttt{OP\_RETURN} payload from the BTC \texttt{scriptpubkey} and perform a hex-aligned comparison against the source hash. This critical step bridges the semantic gap, translating declarative LLM outputs into deterministic byte-level operations (e.g., handling the ASCII decoding of \texttt{OUT:} prefixes).

\begin{figure}[tbp]
\begin{lstlisting}[numbers=none]
def Analyzer(source_tx):
    memo = source_tx["input"]["param"]["memo"]
    m_chain = re.match(r"=:(\w):", memo)
    m_addr  = re.search(r"(bc1[0-9ac-hj-np-z]{11,90})", memo, re.I)
    target_chain = "btc" if m_chain and m_chain.group(1).lower() == "b" else None
    target_addr  = m_addr.group(1).lower() if m_addr else None
    return ...

def Verifier(source_tx, target_tx):
    src = source_tx["hash"].removeprefix("0x").upper()
    out = target_tx["op_returns"][0]["ascii"]
    m = re.search(r"OUT:([0-9A-Fa-f]{64})", out)
    tgt = m.group(1).upper() if m else None
    matched = (src == tgt)
    return ...
\end{lstlisting}
\caption{Python core snippet of the lightweight executable rule operators synthesized by Stage II for THORChain. The \texttt{Analyzer()} function extracts destination side constraints, while the \texttt{Verifier()} function normalizes transaction hashes and extracts payloads from the target \texttt{OP\_RETURN} output to authenticate cross-chain shared evidence.} 
\label{fig:Analyzer_Verifier}
\end{figure}

\begin{figure}[t]
    \centering
    \includegraphics[width=0.95\linewidth]{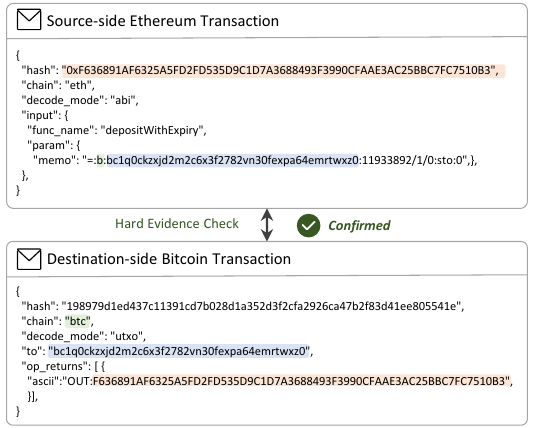}
    \caption{Stage III online correspondence recovery process for THORChain. The system outputs a \textit{Confirmed} forensic state by successfully matching the source Ethereum transaction hash with the payload embedded in the destination Bitcoin \texttt{OP\_RETURN} field.}
    \label{fig:StageIII_Example}
\end{figure}

\subsection{Stage III: Online Correspondence Recovery}
\label{sec:system:stage3}

Stage III operates as the online execution engine of \textsc{XSplicer}. It applies the synthesized rules to live or historical transaction streams, driving the evidence-backed \textit{correspondence} reconstruction.

\textbf{Decoder.} \textsc{XSplicer} first decodes raw input transactions, unrolling low-level traces across heterogeneous ledgers (e.g., EVM, Solana, Bitcoin) into a uniform evidence view. Cross-chain traces manifest differently depending on the execution model: EVM transactions typically embed fields in \textit{calldata}, \textit{event logs}, or internal execution traces; Solana exposes evidence via \textit{instructions}, \textit{inner instructions}, \textit{program logs}, \textit{account metadata}, or \textit{memo} fields; whereas UTXO-based chains rely on \textit{outputs}, \textit{scripts}, or \texttt{OP\_RETURN} payloads.

\textbf{Analyzer \& Collector.} \textsc{XSplicer} invokes the generated \texttt{Analyzer} to extract destination-side constraints from the source transaction. Based on these constraints, it constructs a localized destination-side candidate set ($\mathcal{K}_d$). This collection process utilizes chain-specific adapters designed to rigorously prune the search space from an open-world ledger down to a strictly bounded candidate pool.

\textbf{Verifier.} During candidate evaluation, \textsc{XSplicer} utilizes the \texttt{Verifier} to execute an evidence-stratified model, relying on heuristic soft clues strictly as a fallback when deterministic hard evidence is structurally absent. Rather than forcing a single transaction match, Stage III outputs a rigorous forensic decision state. We define three primary states:
\begin{itemize}[leftmargin=*]\itemsep0.6pt
    \item \textit{Confirmed}: \rev{If exactly one candidate in $\mathcal{K}_d$ satisfies the protocol-level hard-evidence checks, \textsc{XSplicer} reports the correspondence as \textit{Confirmed}.}
    \item \textit{Supported}: If hard evidence is structurally unobservable, yet a candidate exhibits high proximity in soft clues (amount, time, and address) without competitive ambiguity in the candidate set, the system outputs the match as a weighted investigative lead.
    \item \textit{Unresolved}: If the evidence chain is broken or irreconcilable ambiguity exists (e.g., multiple equal-amount candidates lacking hard evidence), the system explicitly flags the state as unresolved, structurally preventing false positives.
\end{itemize}

\textbf{Stage III example output.} As shown in Fig.~\ref{fig:StageIII_Example}, when a new ETH source transaction (\href{https://etherscan.io/tx/0xF636891AF6325A5FD2FD535D9C1D7A3688493F3990CFAAE3AC25BBC7FC7510B3}{\texttt{0xf636...10b3}}) is observed, \textsc{XSplicer} applies the \texttt{Analyzer} to extract the \texttt{memo}:
\begin{quote}
\small\ttfamily =:b:bc1q0ckzxjd2m2c6x3f2782vn30fexpa64emrtw
xz0:11933892/1/0:sto:0
\end{quote}

The \texttt{Analyzer} isolates the chain marker (\texttt{b} $\to$ Bitcoin) and the destination address (\href{https://btcscan.org/address/bc1q0ckzxjd2m2c6x3f2782vn30fexpa64emrtwxz0}{\texttt{bc1q...wxz0}}). Using these constraints, \textsc{XSplicer} dynamically prunes the global BTC ledger to construct a localized \textit{destination-side candidate set}. The system then iterates through this set using the \texttt{Verifier}. Even if the candidate pool is polluted with equal-amount adversarial transactions, the \texttt{Verifier} pinpoints the exact source hash reference deeply embedded within the \texttt{OP\_RETURN} of the true destination transaction (\href{https://btcscan.org/tx/198979d1ed437c11391cd7b028d1a352d3f2cfa2926ca47b2f83d41ee805541e}{\texttt{1989...541e}}):
\begin{quote}
\small\ttfamily OUT:F636891AF6325A5FD2FD535D9C1D7A3688493F3
990CFAAE3AC25BBC7FC7510B3
\end{quote}
This payload perfectly matches the normalized ETH source hash (stripped of the \texttt{0x} prefix and capitalized). \rev{Supported by this hard evidence, \textsc{XSplicer} immediately bypasses probabilistic soft clues and reports the correspondence as \textit{Confirmed}.}

\begin{table*}[t]
\renewcommand{\arraystretch}{1.1}
\centering
\caption{Dataset scale and ground-truth sources.}
\label{tab:dataset}
\resizebox{\textwidth}{!}{
\begin{tabular}{lllllrr}
\toprule
\textbf{Protocol} & \textbf{Dominant Paradigm} & \textbf{Explorer/API for Ground Truth} & \textbf{Documentation Availability} & \textbf{Supported Directions} & \textbf{Collected Pairs} & \textbf{Avg. Time (s)} \\
\midrule
\rowcolor{gray!10} 
Allbridge & Liquidity-/swap-based & \url{https://core.allbridge.io/explorer} & \url{https://docs-core.allbridge.io/} & BSC / ETH / SOL & 14,999 & 339.70 \\
deBridge & Message-based & \url{https://app.debridge.com/orders} & \url{https://github.com/debridge-finance} & BSC / ETH / SOL & 13,828 & 12.53 \\
\rowcolor{gray!10} 
Wormhole & Message-based & \url{https://wormholescan.io/} & \url{https://docs.wormhole.com/wormhole/} & BSC / BTC / ETH / SOL & \rev{3,243} & 396.80 \\
THORChain & Swap-based & \url{https://thorchain.net/dashboard} & \url{https://docs.thorswap.finance/} & BSC / BTC / ETH & 6,000 & 559.14 \\
\rowcolor{gray!10} 
WanBridge & Gateway-/bridge-based & \url{https://www.wanscan.org/} & \url{https://docs.wanchain.org/products/wanbridge} & BSC / ETH / SOL & \rev{9,752} & 2445.13 \\
Orbiter & Routing-/liquidity-based & \url{https://www.orbiter.finance/explore} & \url{https://github.com/Orbiter-Finance} & BSC / ETH / SOL & \rev{5,986} & 600.80 \\
\rowcolor{gray!10} 
Symbiosis & Swap-/routing-based & \url{https://symbiosis.finance/} & \url{https://docs.symbiosis.finance/} & ETH / SOL & 770 & 173.03 \\
\bottomrule
\end{tabular}
}
\end{table*}

\section{Evaluation and Analysis}
\label{sec:evaluation}

\subsection{Experimental Setup}
\label{sec:evaluation:setup}


\rev{We evaluate \textsc{XSplicer} through these research questions:}

\begin{itemize}[leftmargin=*]\itemsep0.6pt
    \item \textbf{RQ1: Overall performance.} Can \textsc{XSplicer} effectively reconstruct cross-chain correspondence across heterogeneous blockchains without relying on bridge-internal privileged information? (Evaluated in Section~\ref{sec:evaluation:Overall})
    
    \item \rev{\textbf{RQ2: Baseline comparison.} How does \textsc{XSplicer} perform compared with existing state-of-the-art cross-chain tracing tools? (Evaluated in Section~\ref{sec:evaluation:comparative})}
    
    \item \textbf{RQ3: Component and robustness analysis.} What are the individual contributions of \textsc{XSplicer}'s core components, and how resilient is the system against artifact ablation and adversarial ambiguity? (Evaluated in Sections~\ref{sec:evaluation:component} and~\ref{sec:evaluation:robustness})
    
    \item \rev{\textbf{RQ4: Practical forensic analysis.} What limitations do official bridge interfaces exhibit, and how effective is \textsc{XSplicer} in post-mortem and adversarial fund-tracing scenarios? (Evaluated in Section~\ref{sec:evaluation:practical})}
\end{itemize}

\subsubsection{Blockchains and Bridge Protocols}

\rev{We evaluate \textsc{XSplicer} across the EVM, Bitcoin, and Solana ecosystems, which span heterogeneous execution models. The EVM setting includes Ethereum (ETH) and Binance Smart Chain (BSC), account-based smart contract platforms with compatible execution structures but distinct ecosystems. Bitcoin (BTC) represents the UTXO model, where native contract event logs are absent and scripting is limited. On Solana (SOL), a high-throughput, instruction-driven non-EVM ledger, we evaluate destination-side constraint recovery under multi-instruction transactions and account-metadata-based candidate collection.}


\subsubsection{Dataset Construction}

\rev{For quantitative correspondence evaluations, labeled transaction pairs come from accessible explorer/API records (RQ1 and transaction-level analyses in RQ3) or a published dataset (RQ2)~\cite{lin2025connector}.} TABLE~\ref{tab:dataset} summarizes the scale of the API-verifiable dataset.
\rev{Public examples used for system construction are disjoint from the evaluation pairs.}

\rev{For practical cases without complete official correspondence records (RQ4), we manually cross-check both \textsc{XSplicer}'s recovered correspondences and its generated hard-evidence rules against raw on-chain events and protocol-specific invariants (Appendix~\ref{app:data:expert-validation}).}


\subsubsection{Evaluation Metrics}
We evaluate \textsc{XSplicer} using these metrics.

\begin{itemize}[leftmargin=*]\itemsep0.6pt
    \item \textbf{Candidate coverage ($\text{Cov.}$):} Quantifies the recall rate of the destination-side intent recovery stage. It is formally defined as the probability that the ground-truth destination transaction, $tx_d^*$, is successfully bounded within the generated local candidate set $\mathcal{K}_d(tx_s)$ across all queries $\mathcal{Q}$: 
    \begin{equation}
        \text{Cov.} = \frac{|\{tx_s \mid tx_d^* \in \mathcal{K}_d(tx_s)\}|}{|\mathcal{Q}|} \nonumber
    \end{equation}

    \item \textbf{Conditional accuracy ($\text{Acc}_{cov}$):} To separate the discriminative capability of the \texttt{Verifier} from data-collection failures in on-chain trace retrieval, we introduce this oracle-assisted metric. The metric assumes an evaluation oracle that guarantees $tx_d^* \in \mathcal{K}_d$, thereby providing an empirical upper bound on verification accuracy. This allows us to evaluate the precision of the \texttt{Verifier} independently of candidate-collection recall.

    \item \textbf{Hard-evidence availability ratio ($\rho_h$):} The proportion of cross-chain queries for which protocol-level deterministic hard evidence, represented by the unified semantic spec, can be reconstructed using only public artifacts.
    
    \item \textbf{Soft-clue ambiguity ratio ($\rho_s$):} Quantifies the risk of false positives when relying only on heuristic soft clues. It is defined as the percentage of candidate sets $\mathcal{K}_d$ that contain at least one non-target noise transaction whose transferred amount falls within a $\pm5\%$ error margin of the source transaction's amount.
    
    \item \rev{\textbf{Latency (Lat.):} The average time required by the online Stage III pipeline to process one source transaction, from ingestion to the final correspondence decision.}
    
\end{itemize}

\subsection{Overall Performance}
\label{sec:evaluation:Overall}
\rev{TABLE~\ref{tab:perf_bridge} and TABLE~\ref{tab:perf_chain} report \textsc{XSplicer}'s performance across seven mainstream bridge protocols and six heterogeneous chain directions, grouped by bridge and chain pair, respectively.}

\begin{table}[t]
\centering
\caption{Overall performance of \textsc{XSplicer} grouped by bridge protocol}
\label{tab:perf_bridge}
\resizebox{\columnwidth}{!}{
\begin{tabular}{lrrrrrr}
\toprule

\textbf{Protocol} & $\textbf{Cov.}$ \textbf{(\%)} & $\textbf{Acc}_{cov}$ \textbf{(\%)} & $\rho_h$ \textbf{(\%)} & $\rho_s$ \textbf{(\%)} & $|\mathcal{K}_d|$ & $\textbf{Lat.}$ \textbf{(s)} \\
\midrule
\rowcolor{gray!10} 
Allbridge & 99.58 & 96.80 & 91.85 & 33.20 & 9.16 & 123.48 \\
deBridge  & 98.60 & 88.12 & 74.67 & 17.03 & 6.00 & 120.42 \\
\rowcolor{gray!10} 
WanBridge & 97.76 & 90.47 & 89.45 & 18.38 & 8.06 & 120.54 \\
THORChain & 93.17 & 93.17 & 92.65 & 7.14  & 2.79 & 46.41  \\
\rowcolor{gray!10} 
Orbiter   & 98.40 & 91.97 & 66.30 & 33.85 & 7.66 & 112.11 \\
Wormhole  & 98.61 & 98.61 & 77.09 & 56.56 & 6.99 & 146.34 \\
\rowcolor{gray!10} 
Symbiosis & 87.92 & 87.92 & 87.79 & 24.28 & 8.33 & 98.68  \\
\bottomrule
\end{tabular}
}
\end{table}

\rev{\textbf{Forensic discriminative power.} TABLE~\ref{tab:perf_bridge} shows that, once the ground-truth transaction is included in the candidate set, Allbridge and THORChain exceed 93\% $\text{Acc}_{cov}$. The protocol invariants reconstructed in Stages I and II therefore provide discriminative evidence within the covered search space. For Wormhole, despite its high soft-clue ambiguity ratio ($\rho_s = 56.56\%$), the hard-evidence-first strategy filters candidates that are similar only in time or amount and achieves 98.61\% $\text{Acc}_{cov}$.}

\rev{\textbf{Failure analysis.} \textsc{XSplicer} may fail to produce a \textit{Confirmed} correspondence for two main reasons. First, candidate collection may fail to include the correct destination transaction, leaving the \texttt{Verifier} with no correct candidate to evaluate. Candidate coverage exceeds 93\% for most protocols, showing that \textsc{XSplicer} usually narrows the search space effectively. Symbiosis is the main exception: its destination settlement is often obscured by intermediate contract executions and routing steps, reducing coverage to 87.92\%. Second, even when the correct transaction is included in the candidate set, \textsc{XSplicer} may lack the protocol-level hard evidence needed to confirm it uniquely. In such cases, the system returns \textit{Supported} or \textit{Unresolved} rather than forcing a \textit{Confirmed} result.}

\begin{insightbox}
    \textit{\textbf{Insight 1:} Cross-chain linkability can remain observable even when bridge-internal mappings are opaque, because parts of the correspondence are preserved in public protocol artifacts. The core of xTCR is to recover destination-side constraints that prune the search space and then use hard evidence, when available ($\rho_h$), to resolve ambiguity introduced by soft clues ($\rho_s$). \textsc{XSplicer} therefore shifts cross-chain tracing from similarity-based heuristics toward evidence-backed forensic reconstruction.}

\end{insightbox}

\begin{table}[t]
\renewcommand{\arraystretch}{1.3}
\centering
\caption{Overall performance of \textsc{XSplicer} by chain pair}
\label{tab:perf_chain}
\resizebox{\columnwidth}{!}{
\begin{tabular}{lrrrrrr}
\toprule
\textbf{Chain Pair} & $\textbf{Cov.}$ \textbf{(\%)} & $\textbf{Acc}_{cov}$ \textbf{(\%)} & $\rho_h$ \textbf{(\%)} & $\rho_s$ \textbf{(\%)} & $|\mathcal{K}_d|$ & $\textbf{Lat.}$ \textbf{(s)} \\
\midrule
\rowcolor{gray!10} 
ETH$\to$SOL & 98.36 & 97.98 & 97.26 & 28.12 & 7.67 & 130.04 \\
ETH$\to$BTC & 93.41 & 93.40 & 92.88 & 9.51  & 2.61 & 42.88  \\
\rowcolor{gray!10} 
BSC$\to$SOL & 98.75 & 94.91 & 94.85 & 23.63 & 8.07 & 135.10 \\
BSC$\to$BTC & 95.82 & 98.37 & 95.12 & 15.09 & 3.32 & 39.67  \\
\rowcolor{gray!10} 
BTC$\to$ETH & 91.93 & 91.68 & 87.76 & 6.85  & 2.39 & 63.83  \\
SOL$\to$ETH & 98.79 & 84.56 & 54.99 & 30.45 & 7.83 & 106.48 \\
\bottomrule
\end{tabular}
}
\end{table}

\textbf{Directional asymmetry in heterogeneous paths.} The chain-pair breakdown (TABLE~\ref{tab:perf_chain}) exposes a pronounced directional asymmetry in cross-chain forensics. For instance, $\text{Acc}_{cov}$ for ETH$\to$SOL transfers reaches 97.98\%, significantly exceeding the reverse SOL$\to$ETH route at 84.56\%. This discrepancy is not an artifact of ledger speed, but rather stems from asymmetric evidence exposure protocols. During SOL$\to$ETH transfers, bridge protocols systematically strip away publicly verifiable identifiers when transitioning from Solana's non-EVM instruction model to the EVM account model. Consequently, the hard-evidence availability ($\rho_h$) plummets to 54.99\%, while soft-clue ambiguity ($\rho_s$) surges to 30.45\%, forcing \textsc{XSplicer} to downgrade its verification strategy to fragile heuristic fallbacks.

\textbf{Computational overhead and real-time viability.} The correlation between candidate pool size ($|\mathcal{C}_d|$) and average latency ($Lat.$) indicates that \textsc{XSplicer}'s processing overhead is dominated by the complexity of bounding the destination search space, rather than native block intervals. 
When targeting Solana (e.g., ETH$\to$SOL, BSC$\to$SOL), high-frequency state updates and program/account-level unrolling inflate both the candidate pool (${\sim}8$ transactions) and processing latency ($130$--$135$s). Conversely, Bitcoin-bound transfers exhibit highly constrained candidate spaces ($|\mathcal{C}_d| < 3.5$) and lower latencies ($40$--$64$s). Once the destination footprint is publicly visible, \textsc{XSplicer} achieves rapid correlation, though total end-to-end discovery remains naturally bounded by Bitcoin's native confirmation delays and bridge payout pacing. Ultimately, \textsc{XSplicer} delivers near real-time monitoring capabilities, bounding even the most complex Solana forensics to a minute-level resolution.

\begin{insightbox}
    \textit{\textbf{Insight 2:} Cross-chain traceability is directional and asymmetric. \textsc{XSplicer} supports low-latency forensic analysis when public invariants tightly bound the destination search space. In contrast, high-throughput models that obscure destination intent can reduce discriminative precision and increase system latency.}

\end{insightbox}
    
\subsection{\rev{Baseline Comparison}}
\label{sec:evaluation:comparative}

\rev{We compare \textsc{XSplicer} with representative EVM-focused cross-chain tracing tools that rely on EVM-specific observability signals:}
\begin{itemize}[leftmargin=*]\itemsep0.6pt
    \item \rev{\textsc{Connector}~\cite{lin2025connector}: Processes EVM traces and event logs using heuristics but relies on explicit, predefined address references and outputs only raw hash pairs.}
    \item \rev{ABCTracer~\cite{lin2025abctracer}: Combines event-log mining, Named Entity Recognition (NER), and information retrieval to relax Connector's heuristics, but remains limited to EVM$\to$EVM.}
\end{itemize}

Given these architectural assumptions, the baselines are not directly applicable to non-EVM models such as Bitcoin or Solana. For a fair comparison, we restrict the evaluation to EVM-to-EVM scenarios and run the baselines under their default configurations. We evaluate the tools on Celer cBridge~\cite{lin2025connector,celer2018white} along two routes: ETH$\to$Polygon (7,296 pairs) and ETH$\to$BSC (600 pairs).

\rev{TABLE~\ref{tab:baseline} shows that \textsc{XSplicer} achieves higher conditional accuracy ($\text{Acc}_{cov}$) in the baselines' natural EVM setting. The hard-evidence availability metric ($\rho_h$) further distinguishes the approaches: \textsc{XSplicer} reconstructs verifiable protocol invariants for evidence-backed correspondence rather than relying primarily on trace/log-pattern heuristics.}


\begin{table}[t]
\renewcommand{\arraystretch}{1.0}
\centering
\caption{Comparative performance of \textsc{XSplicer} against EVM-focused baselines.}
\label{tab:baseline}
\scalebox{0.85}{
\begin{tabular}{lrrrr}
\toprule
\multirow{2}{*}{\textbf{Method}} & \multicolumn{2}{c}{\textbf{ETH $\to$ POLY}} & \multicolumn{2}{c}{\textbf{ETH $\to$ BSC}} \\
\cmidrule(lr){2-3} \cmidrule(lr){4-5}
 & $\textbf{Acc}_{cov}$ \textbf{(\%)} & $\rho_h$ \textbf{(\%)} & $\textbf{Acc}_{cov}$ \textbf{(\%)} & $\rho_h$ \textbf{(\%)} \\
\midrule
Connector        & 95.65 & -- & 95.35 & -- \\
ABCtracer        & 94.38 & -- & 94.08 & -- \\
\textsc{XSplicer} & 100.00 & 100.00 & 99.89 & 99.89 \\
\bottomrule
\end{tabular}
}
\end{table}

\subsection{Component Analysis}
\label{sec:evaluation:component}

\subsubsection{Influence of the LLM Backbone}

\rev{Stage I is the primary phase in which \textsc{XSplicer} explicitly relies on LLMs, so its output quality bounds downstream rule generation and correspondence recovery. We compare foundational models on 70 samples (10 transaction pairs per bridge) using uniform prompts and document slices across three dimensions:}

\begin{itemize}[leftmargin=*]\itemsep0.6pt
    \item \rev{\textbf{Destination-side spec correctness ($\text{F1}_{dst}$):} Accuracy of extracting destination-chain identifiers and address constraints, which affect candidate coverage ($\text{Cov.}$);}
    \item \rev{\textbf{Cross-chain shared spec correctness ($\text{F1}_{cross}$):} Accuracy of recovering cross-chain mapping keys and invariant logic, which affect conditional accuracy ($\text{Acc}_{cov}$);}
    \item \rev{\textbf{Groundedness ($\text{Gnd}$):} Whether the generated spec is anchored to the provided documents and traces, severely penalizing unverified content drawn from the LLM's parametric memory.}
\end{itemize}

\rev{Appendix~\ref{appendix:metrics} gives the detailed formulations. The $\text{F1}$ metrics measure semantic alignment with Gold Specs manually curated by two independent security researchers. TABLE~\ref{tab:llm_backbone} reports the reconstruction results.}

\begin{table}[t]
\renewcommand{\arraystretch}{1.3}
\centering
\caption{Comparative performance of LLM backbones on protocol semantic reconstruction.}
\label{tab:llm_backbone}
\scalebox{0.85}{
\begin{tabular}{lrrr}
\toprule
\textbf{Model} & $\text{F1}_{dst}$ & $\text{F1}_{cross}$ & \textbf{Gnd} \\
\midrule
\rowcolor{gray!10} 
GPT-5.4        & 0.835 & 0.667 & 0.977 \\
DeepSeek-Chat  & 0.744 & 0.538 & 0.924 \\
\rowcolor{gray!10} 
Qwen3-max      & 0.683 & 0.500 & 0.920 \\
Qwen3.6-plus   & 0.766 & 0.573 & 0.903 \\
\rowcolor{gray!10} 
Gemini-3-Pro   & 0.671 & 0.531 & 0.842 \\
\bottomrule
\end{tabular}
}
\end{table}

The results reveal a consistent trend across all models: proficiency in destination-side spec systematically outpaces cross-chain-shared spec. For instance, while GPT-5.4 achieves a robust $\text{F1}_{dst}$ of 0.835 for destination recognition, its performance drops to 0.667 when synthesizing complex pairing logic. This discrepancy highlights a fundamental LLM limitation: parsing routing intent is semantically simpler than deducing execution semantics, which demands rigorous logical reasoning. 
This phenomenon precisely explains the results in Section~\ref{sec:evaluation:Overall}, where \textsc{XSplicer} maintains near-perfect coverage, yet its final accuracy remains naturally bounded by the structural complexity of protocol-specific hard evidence. 
\rev{Overall, GPT-5.4 achieves the highest scores across all three metrics, including an $\text{F1}_{cross}$ score 0.094 higher than that of the runner-up, Qwen3.6-plus.} \textsc{XSplicer} employs GPT-5.4 as the underlying LLM in our experiments. Consequently, employing a state-of-the-art model is critical for robustly capturing protocol invariants, ensuring high-fidelity decision anchors for Stage III.

\begin{insightbox}
    \textit{\textbf{Insight 3:} Semantic recovery quality determines the upper bound of forensic reconstruction. LLMs are useful for extracting routing intent, but they can struggle with complex invariants. We therefore treat LLM outputs as heuristic evidence anchors, and rely on deterministic checks in subsequent stages to finalize forensic correspondence.}
\end{insightbox}

\subsubsection{Functional Utility of Input Artifacts}
\label{sec:component:input}

\rev{Stage I uses protocol documentation and execution traces; we assess their contributions independently.}

\rev{\textbf{Semantic contribution of documentation.} Under the \textit{w/o Docs} ablation, WanBridge BTC$\to$ETH Candidate Coverage drops from 72.60\% to 0\% because critical settlement contract addresses become unavailable. For Wormhole ETH$\to$SOL, hard-evidence availability ($\rho_h$) drops from 99.86\% to zero because cryptographic derivation of its composite identifiers from unstructured VAA metadata requires a textual schema~\cite{wormholescan_api}. Allbridge, which exposes hard evidence in raw logs, remains entirely unaffected.}

\textbf{Structural impact of execution traces.} Evaluating 4,999 Allbridge samples reveals that extraction efficacy relies strictly on structural pattern coverage rather than raw volume. Simulating an adverse environment, a minimum-coverage trace yields a mere 1.66\% $\text{Acc}_{cov}$, requiring 8 traces to recover to 96.54\%. In contrast, a maximum-coverage strategy achieves 86.58\% $\text{Acc}_{cov}$ with a single representative trace. Traces thus function strictly to anchor declarative semantics to concrete ledger architectures. The full results are reported in Appendix~\ref{app:impact_of_examples}.

\begin{insightbox}
    \textit{\textbf{Insight 4:} Cross-chain traceability relies on the semantic complementarity between documentation and execution traces. Documentation provides declarative schemas, while traces ground these abstractions in concrete ledger layouts. Representative traces are more useful than large numbers of structurally redundant samples for finalizing forensic evidence.}
\end{insightbox}

\begin{figure}
    \centering
    \includegraphics[width=0.65\linewidth]{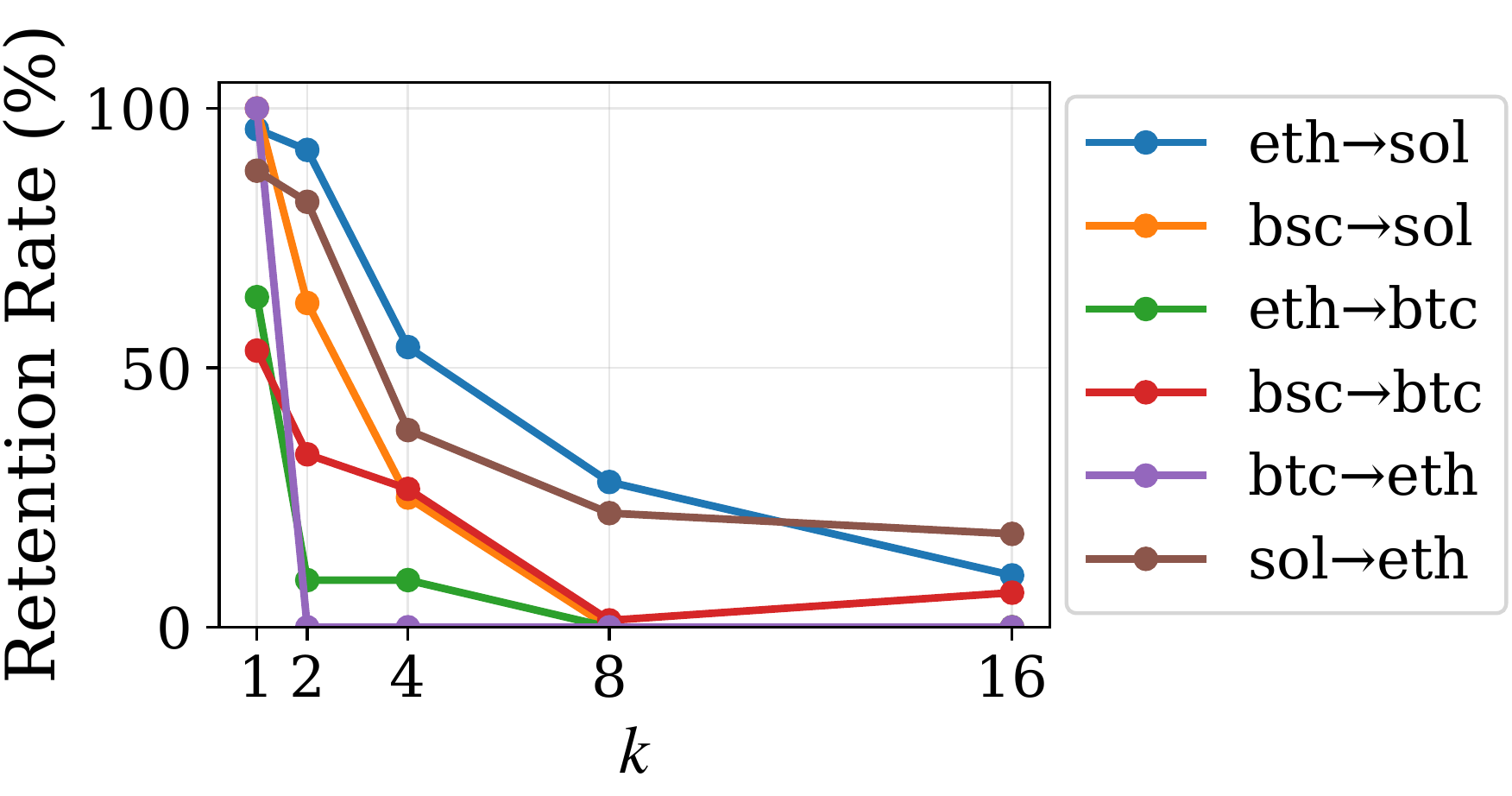}
    \caption{The rapid degradation of retention rates for soft evidence samples across various routes, as the number of injected amount-identical distractors $k$ increases.}
    \label{fig:robustness}
\end{figure}

\subsection{Robustness to Adversarial Ambiguity}
\label{sec:evaluation:robustness}

This part evaluates \textsc{XSplicer}'s resilience against maliciously injected noise. Adversaries can artificially inflate destination-side ambiguity by broadcasting distractor transactions with identical amounts and approximate timestamps.

We evaluate robustness by artificially injecting $k \in \{1,2,4,8,16\}$ amount-identical distractors into the candidate windows of verified samples (randomly selecting up to 50 hard-evidence and 50 soft-evidence cases per route from Section~\ref{sec:evaluation:Overall}). We measure resilience via the \textit{Retention Rate}: the percentage of samples where \textsc{XSplicer} correctly isolates the ground-truth transaction from the polluted candidate pool.

The experiment results show that \textbf{hard-evidence samples maintain a 100\% retention rate} across all routes and $k$ values. \rev{Protocol-level invariants (e.g., sequence numbers or hash references) enabled the verifier to retain the correct correspondence as amount-identical distractors were added.} Conversely, as shown in Fig.~\ref{fig:robustness}, soft-evidence retention rate degrades precipitously as ambiguity scales. The ETH$\to$SOL retention drops from 96\% to 10\% at $k=16$. In inherently ambiguous routes like BTC$\to$ETH (characterized by high $\rho_s$), retention collapses to zero at merely $k=2$. These findings confirm the extreme vulnerability of heuristic matching in adversarial or high-concurrency settings.

\begin{insightbox}
    \textit{\textbf{Insight 5:} Hard-evidence verification accuracy is largely insensitive to candidate-pool size and ambiguity intensity. In contrast, the discriminative power of soft clues degrades rapidly as distractor density increases. This contrast supports protocol-level invariant reconstruction as a practical path toward forensic-grade traceability.}
\end{insightbox}

\subsection{Practical Forensic Analysis}
\label{sec:evaluation:practical}

\rev{This part evaluates the forensic completeness of official bridge interfaces and \textsc{XSplicer}'s utility in post-mortem and adversarial investigations.}

\subsubsection{\rev{Limitations of Official Bridge Interfaces}}
\label{sec:comparison:bridge_interfaces}

\rev{We first examine whether third-party analysts can rely on official bridge explorers or APIs as stable and complete correspondence oracles.} Our empirical analysis of real-world transactions shows that official interfaces exhibit systematic \textbf{failure modes} that limit their forensic completeness. \rev{To assess \textsc{XSplicer} under these blind spots, we manually cross-check its recovered correspondences against public on-chain evidence and protocol-specific invariants.} \rev{In the following three failure modes, official interfaces remain operational but return incomplete correspondence records.}

    \textbf{Failure mode 1: Null results for historical transactions.} Official APIs routinely fail to map historical cross-chain transfers. For instance, within a sample of \textbf{1,988} Orbiter SOL$\to$ETH transactions, the official API silently drops 1,856 records (93.36\%), yielding a mere 6.64\% retrieval rate. Manual inspection confirms this stems from the explorer's UI-centric design, which only indexes the ``Latest Transactions.'' \textsc{XSplicer} successfully reconstructs these severed links using immutable on-chain evidence, demonstrating its critical utility for long-term forensic tracing.

    \textbf{Failure mode 2: Blind spots in one-to-many correspondence.} Although our core formulation focuses on the correspondence between a source transaction $tx_s$ and the final destination-side settlement transaction $tx_d$, real bridge executions do not always produce a single destination-side transaction. Some protocols generate multiple destination-side related transactions $t_{rel}$ around settlement, such as asset splitting or distribution. We refer to this pattern as one-to-many correspondence: one source-chain cross-chain action corresponds to one primary settlement target, or to a set of destination-side related transactions derived from the same protocol execution path. We provide a representative one-to-two case in Appendix~\ref{app:thorchain-one-to-many}. This case is drawn from our analysis of the THORChain attack~\cite{wu2025bridgeguard}, where we identify 17 associated records that are not fully displayed by the official explorer, including transaction \href{https://etherscan.io/tx/0x4f61a1590c4aa2a30ffab1735f5273b72bc1fbb63f90d5faf1b380312913bb55}{\texttt{0x4f61...bb55}}.

    \textbf{Failure mode 3: Omission in multi-step executions.} Official interfaces typically highlight only the final asset settlement, systematically ignoring prerequisite verifications or intermediate routing hops. Across 1,381 Wormhole transfers, \textsc{XSplicer} discovers 51 intermediate \textit{destination-side related transactions} absent from the official API. By matching the cryptographic triplet (\texttt{\_emitter\_chain\_}, \texttt{\_emitter\_address\_}, and \texttt{\_sequence\_}), we verify these transactions as mandatory intermediate steps (e.g., VAA submission\footnote{In the Wormhole protocol, a Verifiable Action Approval (VAA)~\cite{wormholescan_api} is a cryptographically signed message that attests to an event on the source chain, which is required to complete the cross-chain transfer.}). Consequently, \textsc{XSplicer} does not simply mirror official APIs; instead, it pieces together unobservable trace fragments to provide a full-lifecycle execution view.

\subsubsection{\rev{Post-Mortem Forensics after the Multichain Shutdown}}
\rev{By contrast, Multichain's July 2023 shutdown rendered its official tracing services unavailable~\cite{multichain_news}. To evaluate \textsc{XSplicer} in this post-mortem setting, we select two Multichain contracts covering transfers across heterogeneous chains and between EVM-compatible chains.} For the BTC$\rightarrow$ETH direction, we analyze the anyBTC-related contract \href{https://etherscan.io/address/0x51600B0CfF6Bbf79E7767158C41FD15e968Ec404}{\texttt{0x5160...c404}}. For ETH$\rightarrow$BSC/Arbitrum/Polygon directions, we analyze the contract \href{https://etherscan.io/address/0x51600B0CfF6Bbf79E7767158C41FD15e968Ec404}{\texttt{0xBa8D...0705}}.


\rev{The system} \textbf{reconstructs 1,808 EVM-to-EVM and 116 BTC$\to$ETH correspondence pairs}. Two independent security researchers manually verify these recoveries. Notably, the BTC$\to$ETH subset reveals extreme delays up to 144 days between source deposits and destination transactions (e.g., BTC \href{https://btcscan.org/tx/97e3560e0aaf6510f5b1fa69a175ddb280415a3a021346f5de3f225524387ada}{\texttt{97e3...7ada}} $\to$ Ethereum \href{https://etherscan.io/tx/0xbb738d0de98afef28acecec16cf10d914464c7df9182321fa6c9e2d286b34519}{\texttt{0xbb73...4519}}). Such massive temporal gaps would exponentially inflate the candidate space for heuristic algorithms. \rev{This case shows that protocol-level evidence can support correspondence verification even when temporal proximity is uninformative.}


\begin{insightbox}

    \textit{\textbf{Insight 6:} \rev{Official bridge interfaces do not always provide complete cross-chain transaction visibility due to limitations in indexing and service availability. \textsc{XSplicer} complements them by reconstructing protocol-level semantics across heterogeneous chains.}}
\end{insightbox}

\subsubsection{Adversarial Flow Analysis: The Bybit Hacker}
We analyze funds routed through THORChain during the Bybit Hack \cite{bybit_news,trm_bybit_thorchain} to assess reconstruction efficacy in adversarial environments. 
The system accurately isolates two distinct routing topologies:

\textbf{Completed swaps.} \textsc{XSplicer} successfully recovers 754 ETH$\to$BTC correspondence pairs totaling approximately \textbf{105.6} million USD. Cross-referencing against native THORChain logs confirms 100\% destination hash accuracy. \rev{Protocol invariants encoded in the cross-chain \texttt{memo} provide hard evidence linking the corresponding source and destination transactions.} 
We present the details of the top 10 cross-chain trading pairs involved in asset transfers in Appendix~\ref{app:bybit}.

\textbf{Refunded paths.} The analysis identifies 743 ETH$\to$ETH records representing protocol refunds rather than successful transfers (e.g., \href{https://etherscan.io/tx/0x446FF22E1C3737FC8F6918D121139111A61B6CB0D4C630F375BC0500E7EF0EA7}{\texttt{0x446F...0EA7}} $\to$ \href{https://etherscan.io/tx/0x89FADD718BBF0B68B6976BA8AC4E5E6309290CF789CF0D1D98BE7BEC4E09641E}{\texttt{0x89FA...641E}}). We suspect these records represent adversaries' path redirections via THORChain's refund mechanism (which allows custom refund addresses via \texttt{memo}) rather than noise. \rev{\textsc{XSplicer} correctly classifies them as related destination-side transactions ($tx_{rel}$), distinct from successful settlements ($tx_d$).}

\rev{Overall, the evaluation shows that \textsc{XSplicer} can reconstruct cross-chain correspondence from public artifacts without requiring bridge-internal mappings, including in post-mortem settings where official services are unavailable.} Its explicit modeling of settlement, intermediate execution, refund, and return states further reduces false-positive links during adversarial fund tracing.

\section{Discussion}


\rev{\textbf{Applicability boundaries.} Because its effectiveness depends on public protocol invariants, \textsc{XSplicer} applies to decentralized bridges and contract-driven DeFi flows whose ledgers retain explicit or implicit destination intents and shared evidence. It cannot reconstruct correspondence for custodial bridges with private-backend mappings or anonymity protocols that remove cross-chain identifiers via zero-knowledge proofs. Such cases return \textit{unresolved} rather than an unevidenced correlation.}

\rev{\textbf{Evidence boundaries.} Without protocol-level hard evidence, \textsc{XSplicer} outputs a weighted soft-clue lead only when a single candidate matches both time and amount with extremely high confidence. Multiple matches or excessive noise yield \textit{unresolved}. Appendix~\ref{app:ethical} discusses the corresponding ethical considerations.}

\rev{\textbf{Functional boundaries.} \textsc{XSplicer} reconstructs transaction-level correspondence, not end-to-end multi-hop entity recognition. Its evidence chains supply foundational metadata for address clustering and illicit-flow analysis, but entity attribution and legal conclusions require off-chain intelligence and manual investigation.}

\rev{\textbf{Extensibility.} Onboarding a new bridge only requires collecting public protocol documentation and a transaction example. XSPLICER then automatically generates the corresponding Analyzer and Verifier without manually implementing bridge-specific rules.}

\rev{\textbf{Future work.} We plan to automate invariant extraction from bytecode to reduce documentation dependence, improve autonomy, and extend correspondence reconstruction to mixing bridges~\cite{wu2026TGweaver}.}

\section{Related Work}
\label{sec:related}


\textbf{Cross-chain Interoperability and Bridge Security.}
Multi-chain architectures have made cross-chain bridges core interoperability infrastructure~\cite{hu2024jigsaw,augusto2024sok}. Recent studies examine their architectures, transaction patterns, and economic effects~\cite{huang2024stargate,cao2026price}. Because bridges hold large assets and implement complex cross-chain logic, their security has received sustained attention. Prior work systematizes security and privacy issues in interoperability protocols, including attack surfaces, architectural flaws, and failure modes~\cite{zhang2024bridgeSecurity,augusto2024sok}. Other studies propose defenses, such as malicious transaction detection from known attack patterns~\cite{zhang2022xscope,wu2025bridgeguard}, fine-grained static analysis for contract vulnerabilities~\cite{liao2024smartaxe}, and runtime monitoring of bridge business logic~\cite{eshghie2024highguard}. These works focus on bridge vulnerabilities and defenses. In contrast, we study how a third-party analyst can recover cross-chain transaction correspondence under incomplete observability and without privileged bridge interfaces.


\textbf{Multi-chain Transaction Forensics and Tracing.}
Blockchain forensics has traditionally focused on single-chain fund flows, using transaction features, heuristics, and graph mining to characterize laundering and asset-transfer behavior~\cite{wu2024assetflows}. Surveys further organize these tracking techniques as blockchain forensics evolves~\cite{kumar2025surveyTracing}. Some work extends beyond a single chain by exploiting address consistency across EVM-compatible chains~\cite{yan2025evmPolygon} or tracing flows across cryptocurrency ledgers~\cite{yousaf2019tracing}. Recent systems address cross-chain DeFi transfers by pruning semantic search spaces with large language models~\cite{liang2025connex} or constructing logical relations from multi-chain data to detect anomalies~\cite{augusto2025xchainwatcher}. These studies provide useful foundations for cross-chain tracing. Our work differs in its problem definition and evidence model: rather than matching transactions in homogeneous settings or specific protocols~\cite{lin2025connector,lin2025abctracer}, we recover evidence-backed correspondence across heterogeneous environments, including EVM, Bitcoin, and Solana.

\section{Conclusion}

The proliferation of heterogeneous bridges has precipitated a severe observability crisis, with only 16.79\% of 131 surveyed bridge protocols providing verifiable tracking. \rev{To address this, we present \textsc{XSplicer}, a protocol-evidence compilation system for cross-chain correspondence reconstruction in a non-privileged third-party setting.}
\rev{\textsc{XSplicer} synthesizes public protocol artifacts into a unified semantic specification, compiles the specification into executable analyzers and verifiers, and produces evidence-linked forensic decisions.}
\rev{Across seven evaluated bridge protocols spanning EVM, Solana, and Bitcoin, \textsc{XSplicer} achieves 87.92\%--98.61\% oracle-assisted conditional accuracy.} 
\rev{Its hard-evidence verification supports post-mortem traceability even when official services are unavailable and remains resilient to adversarial noise. The Multichain and Bybit case studies further demonstrate its practical forensic value in reconstructing historical cross-chain correspondences and tracing illicit fund flows.}

\newpage

\bibliographystyle{plainurl}
\bibliography{usenixsecurity2027}

@inproceedings {yousaf2019tracing,
author = {Haaroon Yousaf and George Kappos and Sarah Meiklejohn},
title = {Tracing transactions across cryptocurrency ledgers},
booktitle = {USENIX Security Symposium},
year = {2019},
pages = {837--850},
month = {Aug.},
doi = {10.5555/3361338.3361396}
}

@article{belchior2021survey,
  author    = {Rafael Belchior and Andr{\'e} Vasconcelos and S{\'e}rgio Guerreiro and Miguel Correia},
  title     = {A Survey on Blockchain Interoperability: Past, Present, and Future Trends},
  journal   = {ACM Computing Surveys},
  volume    = {54},
  number    = {8},
  articleno = {168},
  year      = {2022},
  doi       = {10.1145/3471140}
}

@inproceedings{zamyatin2021sok,
  author    = {Alexei Zamyatin and Mustafa Al-Bassam and Dionysis Zindros and Eleftherios Kokoris-Kogias and Pedro Moreno-Sanchez and Aggelos Kiayias and William J. Knottenbelt},
  title     = {{SoK}: Communication Across Distributed Ledgers},
  booktitle = {Financial Cryptography and Data Security},
  series    = {Lecture Notes in Computer Science},
  volume    = {12675},
  pages     = {3--36},
  year      = {2021},
  publisher = {Springer Nature},
  doi       = {10.1007/978-3-662-64331-0_1}
}

@article{li2025interoperability,
  author  = {Wenqing Li and Zhenguang Liu and Jianhai Chen and Zhe Liu and Qinming He},
  title   = {Towards Blockchain Interoperability: A Comprehensive Survey on Cross-Chain Solutions},
  journal = {Blockchain: Research and Applications},
  volume  = {6},
  number  = {3},
  pages   = {100286},
  year    = {2025},
  doi     = {10.1016/j.bcra.2025.100286}
}

@article{han2023survey,
  author  = {Panpan Han and Zheng Yan and Wenxiu Ding and Shufan Fei and Zhiguo Wan},
  title   = {A Survey on Cross-Chain Technologies},
  journal = {Distributed Ledger Technologies: Research and Practice},
  volume  = {2},
  number  = {2},
  pages   = {1--30},
  year    = {2023},
  doi     = {10.1145/3573896}
}

@misc{defillama_bridges,
  author       = {{DefiLlama}},
  title        = {DefiLlama Bridges Dashboard},
  howpublished = {\url{https://defillama.com/bridges}},
  year         = {2026},
  note         = {Accessed: 2026-01-13}
}

@article{gramlich2023multivocal,
  author  = {Vincent Gramlich and Tobias Guggenberger and Marc Principato and Benjamin Schellinger and Nils Urbach},
  title   = {A Multivocal Literature Review of Decentralized Finance: Current Knowledge and Future Research Avenues},
  journal = {Electronic Markets},
  volume  = {33},
  number  = {1},
  pages   = {11},
  year    = {2023},
  doi     = {10.1007/s12525-023-00637-4}
}

@article{kitzler2023defi,
  author  = {Stefan Kitzler and Friedhelm Victor and Pietro Saggese and Bernhard Haslhofer},
  title   = {Disentangling Decentralized Finance ({DeFi}) Compositions},
  journal = {ACM Transactions on the Web},
  volume  = {17},
  number  = {2},
  articleno = {10},
  pages   = {1--26},
  year    = {2023},
  doi     = {10.1145/3532857}
}

@misc{multichain_news,
  author       = {{CoinDesk}},
  title        = {{Recently Exploited Crypto Bridge Shuts, Says China Detained CEO and His Sister}},
  howpublished = {\url{https://www.coindesk.com/business/2023/07/14/crypto-bridging-protocol-multichain-ceases-operations}},
  year         = {2023},
  note         = {Accessed: 2026-08-24}
}

@misc{chainspot_list,
  title = {{Chainspot: Bridges}},
  author = {{Chainspot}},
  howpublished = {\url{https://chainspot.io/portal/bridges}},
  year = {2024},
  note = {Accessed: 2026-08-24}
}

@misc{across_api,
  title = {{Across Developer Documentation: API Reference}},
  author = {{Across Protocol}},
  howpublished = {\url{https://docs.across.to/reference/api-reference}},
  year = {2026},
  note = {Accessed: 2026-08-24}
}

@misc{wormholescan_api,
  title = {{Wormhole Documentation: Fetch a Signed VAA}},
  author = {{Wormhole Foundation}},
  howpublished = {\url{https://wormhole.com/docs/products/token-transfers/wrapped-token-transfers/guides/fetch-signed-vaa/}},
  year = {2026},
  note = {Accessed: 2026-08-24}
}

@misc{thorchain_memos,
  author       = {{THORChain}},
  title        = {Transaction Memos},
  howpublished = {\url{https://dev.thorchain.org/concepts/memos.html}},
  year         = {2026},
  note         = {Accessed: 2026-05-04}
}

@misc{thorchain_sending_transactions,
  author       = {{THORChain}},
  title        = {Sending Transactions},
  howpublished = {\url{https://dev.thorchain.org/concepts/sending-transactions.html}},
  year         = {2026},
  note         = {Accessed: 2026-05-04}
}

@article{lin2025connector,
  author  = {Dan Lin and Jiajing Wu and Yuxin Su and Ziye Zheng and Yuhong Nan and Qinnan Zhang and Bowen Song and Zibin Zheng},
  title   = {Connector: Enhancing the Traceability of Decentralized Bridge Applications via Automatic Cross-Chain Transaction Association},
  journal = {IEEE Transactions on Information Forensics and Security},
  volume  = {20},
  pages   = {7588--7601},
  year    = {2025},
  doi     = {10.1109/TIFS.2025.3588249}
}

@article{lin2025abctracer,
  author  = {Dan Lin and Ziye Zheng and Jiajing Wu and Jingjing Yang and Kaixin Lin and Huan Xiao and Bowen Song and Zibin Zheng},
  title   = {Track and Trace: Automatically Uncovering Cross-Chain Transactions in the Multi-Blockchain Ecosystems},
  journal = {IEEE Transactions on Services Computing},
  volume  = {18},
  number  = {6},
  pages   = {4291--4303},
  year    = {2025},
  doi     = {10.1109/TSC.2025.3618729}
}

@misc{elliptic2025,
  author       = {{Elliptic}},
  title        = {The State of Cross-chain Crime 2025},
  howpublished = {\url{https://www.elliptic.co/resources/the-state-of-cross-chain-crime-2025}},
  year         = {2025},
  note         = {Accessed: 2026-05-04}
}

@misc{layerzero_value_transfer_api,
  author       = {{LayerZero Labs}},
  title        = {Value Transfer API},
  howpublished = {\url{https://docs.layerzero.network/v2/developers/value-transfer-api/start}},
  year         = {2026},
  note         = {Accessed: 2026-05-04}
}

@misc{debridge_lifecycle,
  author       = {{deBridge}},
  title        = {Cross-Chain Call Lifecycle},
  howpublished = {\url{https://docs.debridge.com/dmp-details/dev-guides/cross-chain-call-lifecycle}},
  year         = {2026},
  note         = {Accessed: 2026-05-04}
}

@misc{debridge_dln_overview,
  author       = {{deBridge}},
  title        = {deBridge Liquidity Network Overview},
  howpublished = {\url{https://docs.debridge.com/dln-details/overview/introduction}},
  year         = {2026},
  note         = {Accessed: 2026-05-04}
}

@misc{debridge_dln_contracts,
  author       = {{deBridge}},
  title        = {deBridge Liquidity Network Deployed Contracts},
  howpublished = {\url{https://docs.debridge.com/dln-details/overview/deployed-contracts}},
  year         = {2026},
  note         = {Accessed: 2026-05-04}
}

@misc{ethereum_transactions,
  author       = {{Ethereum Foundation}},
  title        = {Ethereum Transactions},
  howpublished = {\url{https://ethereum.org/developers/docs/transactions/}},
  year         = {2026},
  note         = {Accessed: 2026-05-04}
}

@misc{bitcoin_transactions,
  author       = {{Bitcoin Developer Documentation}},
  title        = {Transactions},
  howpublished = {\url{https://developer.bitcoin.org/devguide/transactions.html}},
  year         = {2026},
  note         = {Accessed: 2026-05-04}
}

@misc{solana_transactions,
  author       = {{Solana Foundation}},
  title        = {Transactions},
  howpublished = {\url{https://solana.com/docs/core/transactions}},
  year         = {2026},
  note         = {Accessed: 2026-05-04}
}

@misc{stargate_api_deprecated,
  author       = {{Stargate Finance}},
  title        = {Stargate API},
  howpublished = {\url{https://docs.stargate.finance/developers/api-docs/overview}},
  year         = {2025},
  note         = {Accessed: 2026-05-04}
}

@inproceedings{wu2025bridgeguard,
  author    = {Jiajing Wu and Kaixin Lin and Dan Lin and Bozhao Zhang and Zhiying Wu and Jianzhong Su},
  title     = {Safeguarding Blockchain Ecosystem: Understanding and Detecting Attack Transactions on Cross-chain Bridges},
  booktitle = {Proceedings of the ACM Web Conference 2025},
  pages     = {4902--4912},
  year      = {2025},
  doi       = {10.1145/3696410.3714604}
}

@misc{bybit_news,
  author       = {{Bybit}},
  title        = {Bybit Security Incident: Timeline of Events and FAQs},
  howpublished = {\url{https://learn.bybit.com/en/this-week-in-bybit/bybit-security-incident-timeline}},
  year         = {2025},
  note         = {Accessed: 2026-05-04}
}

@misc{trm_bybit_thorchain,
  author       = {{TRM Labs}},
  title        = {Bybit Hack Update: North Korea Moves to Next Stage of Laundering},
  howpublished = {\url{https://www.trmlabs.com/resources/blog/bybit-hack-update-north-korea-moves-to-next-stage-of-laundering}},
  year         = {2025},
  note         = {Accessed: 2026-05-04}
}

@article{hu2024jigsaw,
  author     = {Xiaohui Hu and Hang Feng and Pengcheng Xia and Gareth Tyson and Lei Wu and Yajin Zhou and Haoyu Wang},
  title      = {Piecing Together the Jigsaw Puzzle of Transactions on Heterogeneous Blockchain Networks},
  journal    = {Proceedings of the ACM on Measurement and Analysis of Computing Systems},
  volume     = {8},
  number     = {3},
  articleno  = {42},
  pages      = {1--27},
  year       = {2024},
  doi        = {10.1145/3700424},
  url        = {https://doi.org/10.1145/3700424}
}

@inproceedings{augusto2024sok,
  author    = {Andr{\'e} Augusto and Rafael Belchior and Miguel Correia and Andr{\'e} Vasconcelos and Luyao Zhang and Thomas Hardjono},
  title     = {{SoK}: Security and Privacy of Blockchain Interoperability},
  booktitle = {2024 IEEE Symposium on Security and Privacy (SP)},
  pages     = {3840--3865},
  year      = {2024},
  publisher = {IEEE},
  doi       = {10.1109/SP54263.2024.00255}
}

@inproceedings{huang2024stargate,
  author    = {Chuanshan Huang and Tao Yan and Claudio J. Tessone},
  title     = {Seamlessly Transferring Assets through Layer-0 Bridges: An Empirical Analysis of Stargate Bridge's Architecture and Dynamics},
  booktitle = {Companion Proceedings of the ACM Web Conference 2024},
  pages     = {1776--1784},
  year      = {2024},
  doi       = {10.1145/3589335.3651964},
}

@inproceedings{cao2026price,
  author = {Cao, Yiyue and Zheng, Mingzhe and Cong, Lin William and Li, Siguang and Wang, Xuechao},
  title = {The Price of Interoperability: Exploring Cross-Chain Bridges and Their Economic Consequences},
  year = {2026},
  booktitle = {Proceedings of the ACM on Measurement and Analysis of Computing Systems},
  doi = {10.1145/3805650},
}

@inproceedings{zhang2024bridgeSecurity,
  author    = {Mengya Zhang and Xiaokuan Zhang and Yinqian Zhang and Zhiqiang Lin},
  title     = {Security of Cross-chain Bridges: Attack Surfaces, Defenses, and Open Problems},
  booktitle = {Proceedings of the International Symposium on Research in Attacks, Intrusions and Defenses},
  pages     = {298--316},
  year      = {2024},
  doi       = {10.1145/3678890.3678894},
}

@inproceedings{zhang2022xscope,
  author    = {Jiashuo Zhang and Jianbo Gao and Yue Li and Ziming Chen and Zhi Guan and Zhong Chen},
  title     = {Xscope: Hunting for Cross-Chain Bridge Attacks},
  booktitle = {Proceedings of the IEEE/ACM International Conference on Automated Software Engineering},
  articleno = {171},
  pages     = {1--4},
  year      = {2022},
  doi       = {10.1145/3551349.3559520},
}

@article{liao2024smartaxe,
  author     = {Zeqin Liao and Yuhong Nan and Henglong Liang and Sicheng Hao and Juan Zhai and Jiajing Wu and Zibin Zheng},
  title      = {SmartAxe: Detecting Cross-Chain Vulnerabilities in Bridge Smart Contracts via Fine-Grained Static Analysis},
  journal    = {Proceedings of the ACM on Software Engineering},
  pages      = {249--270},
  year       = {2024},
  doi        = {10.1145/3643738},
}

@inproceedings{eshghie2024highguard,
  author    = {Mojtaba Eshghie and Cyrille Artho and Hans Stammler and Wolfgang Ahrendt and Thomas Hildebrandt and Gerardo Schneider},
  title     = {HighGuard: Cross-Chain Business Logic Monitoring of Smart Contracts},
  booktitle = {Proceedings of the IEEE/ACM International Conference on Automated Software Engineering},
  pages     = {2378--2381},
  year      = {2024},
  doi       = {10.1145/3691620.3695356},
}

@article{wu2024assetflows,
  author  = {Jiajing Wu and Dan Lin and Qishuang Fu and Shuo Yang and Ting Chen and Zibin Zheng and Bowen Song},
  title   = {Toward Understanding Asset Flows in Crypto Money Laundering Through the Lenses of Ethereum Heists},
  journal = {IEEE Transactions on Information Forensics and Security},
  volume  = {19},
  pages   = {1994--2009},
  year    = {2024},
  doi     = {10.1109/TIFS.2023.3346276}
}

@misc{kumar2025surveyTracing,
  author       = {Ayush Kumar and Vrizlynn L. L. Thing},
  title        = {A Survey of Transaction Tracing Techniques for Blockchain Systems},
  year         = {2025},
  howpublished = {arXiv preprint arXiv:2510.09624},
  url          = {https://arxiv.org/abs/2510.09624}
}

@article{yan2025evmPolygon,
  author  = {Tao Yan and Chuanshan Huang and Claudio J. Tessone},
  title   = {Tracing Cross-Chain Transactions Between {EVM}-Based Blockchains: An Analysis of {Ethereum}-{Polygon} Bridges},
  journal = {Ledger},
  volume  = {10},
  year    = {2025},
  pages = {113--134},
  doi     = {10.5195/ledger.2025.433}
}

@misc{liang2025connex,
  author       = {Hanzhong Liang and Yue Duan and Xing Su and Xiao Li and Yating Liu and Yulong Tian and Fengyuan Xu and Sheng Zhong},
  title        = {ConneX: Automatically Resolving Transaction Opacity of Cross-Chain Bridges for Security Analysis},
  year         = {2025},
  howpublished = {arXiv preprint arXiv:2511.01393},
  url          = {https://arxiv.org/abs/2511.01393}
}

@inproceedings{augusto2025xchainwatcher,
  author    = {Andr{\'e} Augusto and Rafael Belchior and Jonas Pfannschmidt and Andr{\'e} Vasconcelos and Miguel Correia},
  title     = {XChainWatcher: Identifying Anomalies in Cross-Chain Bridges},
  booktitle = {Proceedings of the International Middleware Conference},
  pages     = {413--426},
  year      = {2025},
  doi       = {10.1145/3721462.3770781},
  url       = {https://doi.org/10.1145/3721462.3770781}
}

@misc{thorchain_dev_docs,
  title        = {{THORChain Dev Docs}},
  author       = {{THORChain}},
  howpublished = {\url{https://dev.thorchain.org/}},
  note         = {Accessed: 2026-05-07}
}

@inproceedings{brown2020language,
  title={Language Models are Few-Shot Learners},
  author={Brown, Tom B. and Mann, Benjamin and Ryder, Nick and Subbiah, Melanie and Kaplan, Jared and Dhariwal, Prafulla and Neelakantan, Arvind and Shyam, Pranav and Sastry, Girish and Askell, Amanda and others},
  booktitle={Advances in Neural Information Processing Systems},
  volume={33},
  pages={1877--1901},
  year={2020}
}

@inproceedings{wei2022chain,
  title={Chain-of-Thought Prompting Elicits Reasoning in Large Language Models},
  author={Wei, Jason and Wang, Xuezhi and Schuurmans, Dale and Bosma, Maarten and Ichter, Brian and Xia, Fei and Chi, Ed and Le, Quoc and Zhou, Denny},
  booktitle={Proceedings of International Conference on Neural Information Processing Systems},
  volume={35},
  pages={24824--24837},
  year={2022}
}

@inproceedings{lewis2020retrieval,
  title={Retrieval-Augmented Generation for Knowledge-Intensive NLP Tasks},
  author={Lewis, Patrick and Perez, Ethan and Piktus, Aleksandra and Petroni, Fabio and Karpukhin, Vladimir and Goyal, Naman and K{\"u}ttler, Heinrich and Lewis, Mike and Yih, Wen-tau and Rockt{\"a}schel, Tim and Riedel, Sebastian and Kiela, Douwe},
  booktitle={Advances in Neural Information Processing Systems},
  volume={33},
  pages={9459--9474},
  year={2020},
  url={https://proceedings.neurips.cc/paper_files/paper/2020/file/6b493230205f780e1bc26945df7481e5-Paper.pdf}
}

@inproceedings{yao2023react,
  title={ReAct: Synergizing Reasoning and Acting in Language Models},
  author={Yao, Shunyu and Zhao, Jeffrey and Yu, Dian and Du, Nan and Shafran, Izhak and Narasimhan, Karthik and Cao, Yuan},
  booktitle={International Conference on Learning Representations},
  year={2023},
  url={https://arxiv.org/pdf/2210.03629}
}

@inproceedings{ wu2026TGweaver,
author = {Wu, Fajie and Wu, Jiajing and Wu, Zhiying and Chen, Jun and Wang, Tao and He, Longjian and Song, Bowen and Wang, Weiqiang},
title = {TGweaver: Synthesizing Transaction Graphs for De-anonymization Analysis},
booktitle = {Proceedings of the ACM Web Conference},
year = {2026},
doi = {10.1145/3774904.3792318},
pages = {2835-2845}
}

@online{ celer2018white,
  author =       {{ScaleSphere Foundation Ltd. }},
  year =         {2018},
  month=         {Jun.},
  title =        {Celer Network: {B}ring internet scale to every blockchain},
  howpublished = {\url{https://celer.network/doc/celernetwork-whitepaper.pdf}}
}


\appendix

\section{Open Science}

The core \textsc{XSplicer} implementation, benchmark subsets, and illicit-flow traces will be made publicly available upon publication to support reproducibility and further research.
To mitigate privacy risks and dual-use concerns, we withhold large-scale transaction dumps containing benign user addresses and operational scripts facilitating evasion tactics. 
Following responsible disclosure principles, we have proactively \textbf{emailed the respective development teams} regarding the systematic oracle failures and observability blind spots identified in Section~\ref{sec:comparison:bridge_interfaces}.



\section{Ethical Considerations}
\label{app:ethical}

Our cross-chain forensic methodology strictly complies with Menlo Report guidelines.

\textbf{Data privacy.} All utilized artifacts derive exclusively from public repositories. We strictly eschew private records including centralized KYC data, internal backend mappings, and proprietary relayer states. While \textsc{XSplicer} enhances cross-chain linkability, its scope remains strictly confined to logical transaction correspondence. We perform zero natural person de-anonymization, address ownership inference, or legal attribution. All benign user addresses remain fully anonymized.

\textbf{Non-intrusive methodology.} We employ purely passive observation. The methodology involves zero active probing, adversarial transaction submissions, or stress testing against live bridge protocols. Official APIs serve exclusively as evaluation oracles rather than system inputs. This completely offline paradigm guarantees zero interference with production services.

\textbf{False positive mitigation.} Erroneous correlations severely undermine compliance tracking. \textsc{XSplicer} mitigates this via an evidence-driven architecture outputting three strict decision states: hard-evidence confirmation, soft-clue support, and unresolved. This tripartite model structurally prevents probabilistic guesswork in forensic conclusions.

\textbf{Responsible disclosure.} We recognize the dual-use potential of exposing protocol tracing mechanics. We deliberately omit operational scripts facilitating evasion tactics. 

Following responsible disclosure protocols, we proactively emailed the respective bridge developers regarding the systematic API blind spots identified in Section~\ref{sec:comparison:bridge_interfaces}.

\section{\rev{Implications and Recommendations}}

\rev{Our findings have implications for three stakeholder groups. Bridge developers should document cross-chain invariants and retain public evidence anchors to support independent audits and post-mortem reconstruction. Forensic analysts and compliance systems can use protocol-level evidence to complement time-and-amount heuristics in heterogeneous cross-chain investigations. Finally, users should assess a bridge's privacy properties from its protocol design and on-chain evidence rather than from the opacity of its interface.}

\section{Cross-Chain Bridge Survey}
\label{app:data:bridge-survey}

To comprehensively substantiate the pervasive ``observability crisis'' discussed in Section~\ref{sec:background:observability}, TABLE~\ref{tab:bridge-survey-two-panel} provides the complete empirical survey results of 131 cross-chain bridges and interoperability protocols. These targets were systematically sourced from the cross-chain aggregator Chainspot~\cite{chainspot_list}. Utilizing an industry-recognized aggregator ensures that our empirical sample is highly representative of the actively utilized DeFi ecosystem, strictly mitigating subjective selection bias.
It is crucial to emphasize that the mere existence of an official explorer or a developer API does not equate to forensic traceability. 

In this survey, a feature is positively marked (\ding{51}) \textit{strictly} if it provides a stable, public, and verifiable mapping from a source transaction hash ($tx_s$) to its exact destination transaction hash ($tx_d$). We apply a hierarchical strategy: features failing this standard are marked inadequate (\ding{55}). Conversely, if foundational infrastructure is absent, dependent checks are omitted and marked with a dash (--).

The table systematically details each protocol's official documentation links, block explorer accessibility, support for heterogeneous execution environments (EVM, Bitcoin, Solana), and API mapping capabilities. The extensive prevalence of missing or inadequate tracking interfaces across these 131 protocols constitutes the empirical foundation for our assertion that \rev{third-party analysts frequently lack stable and complete official source-to-destination mappings, motivating} the evidence-backed reconstruction approach of \textsc{XSplicer}.

\section{Cross-chain Transactions in Bybit Hack Laundering}
\label{app:bybit}
To further illustrate the real-world tracking capabilities of \textsc{XSplicer} in high-value adversarial scenarios, TABLE~\ref{tab:tx_tracking} details the top 10 cross-chain transaction pairs recovered during the Bybit hack analysis. By deterministically matching the source transactions on Ethereum with their corresponding destination settlements on Bitcoin, the table demonstrates the system's ability to reconstruct the precise illicit fund flows and verify the exact transfer amounts without relying on official bridge interfaces.

\begin{table}[t]
\centering
\caption{Top 10 high-value cross-chain transaction pairs reconstructed during the Bybit hack money laundering.}
\label{tab:tx_tracking}
\resizebox{\columnwidth}{!}{
\renewcommand{\arraystretch}{1}
\begin{tabular}{cclclcc}
\toprule
\textbf{\#} & \textbf{Source Asset} & \textbf{Source Tx} & \textbf{Destination Asset} & \textbf{Destination Tx} & \textbf{BTC} \\
\midrule
\rowcolor{gray!10} 
1  & ETH.ETH & \href{https://etherscan.io/tx/0x219FE073729A92D534B9789A90795004191350A6E293BF6A0EE1A9738C81A76E}{0x219F\dots A76E} & BTC.BTC & \href{https://btcscan.org/tx/206B855455175F24B1F5E2061DF3607BA4CA5191FD895D76D0E354A0AC762DD0}{206B\dots 2DD0} & 7.5071 \\
2  & ETH.ETH & \href{https://etherscan.io/tx/0x5B49734F1D5310A7B05DDDE884A212183D3AB01041EC61952EBC00322E97017A}{0x5B49\dots 017A} & BTC.BTC & \href{https://btcscan.org/tx/73A82F4423A8AB5AB089D1B5E22582073AB727207297791515191AFC9E2827FB}{73A8\dots 27FB} & 7.3861 \\
\rowcolor{gray!10} 
3  & ETH.ETH & \href{https://etherscan.io/tx/D48D05D2021524E57007A2527A193AF115DC24E1CA80F884AC4F1711F2279E75}{0xD48D\dots 9E75} & BTC.BTC & \href{https://btcscan.org/tx/DB95DD37BE308BE57DB3D9A2C4A1C97E8E11C562C444D86A80D44E80ADBCDBDE}{DB957\dots DBDE} & 7.286  \\
4  & ETH.ETH & \href{https://etherscan.io/tx/0x628466327F35C4708E00FF56CB49F2671C64AC13DAD3E013021D7C800F88F982}{0x6284\dots F982} & BTC.BTC & \href{https://btcscan.org/tx/D51F036E16FC50E49849955FDA4016FB51F03BA9D32EE55B295BDF67D572114C}{D51F\dots 114C} & 7.2355 \\
\rowcolor{gray!10} 
5  & ETH.ETH & \href{https://etherscan.io/tx/0xBF877C7F95D6F79D0452FDA8370E9600AA6337E4A7E9A28774C23C9A46B90D41}{0xBF87\dots 0D41} & BTC.BTC & \href{https://btcscan.org/tx/4EBB1E2C2BC172C52ABCBCC6059B32DF092E097708759A4EDFF6716E3CE5ED67}{4EBB\dots ED67} & 7.2041 \\
6  & ETH.ETH & \href{https://etherscan.io/tx/0x7468AE7C8372BB271AF3D88F5214699198936210FBF06CCB30375978BA5A5FCB}{0x7468\dots 5FCB} & BTC.BTC & \href{https://btcscan.org/tx/12A87E20AFADCF4D4F19DB517AD9B471F58C25A8EAA678D0CFC5696859CD33D7}{12A8\dots 33D7} & 7.1175 \\
\rowcolor{gray!10} 
7  & ETH.ETH & \href{https://etherscan.io/tx/0xD4230A6A5277869FAFEEFE6DAFA863810354D8E1FC7DA6814D6DA70FB7FDD43F}{0xD423\dots D43F} & BTC.BTC & \href{https://btcscan.org/tx/97F7A15F8F303AA38E8C435B3A7BA1F38E12A217C1EBBA17ABB677B073FC4AE7}{97F7\dots 4AE7} & 7.037  \\
8  & ETH.ETH & \href{https://etherscan.io/tx/0xF10D05185903FE0ED2709080FBDC6C37E68DC5C994AB34707DD69B47142A262F}{0xF10D\dots 262F} & BTC.BTC & \href{https://btcscan.org/tx/7DC349D7F47BDB33E52EF9C71FAC06189D50BCE5BE66B408A47C853691B87E81}{7DC3\dots 7E81} & 6.9575 \\
\rowcolor{gray!10} 
9  & ETH.ETH & \href{https://etherscan.io/tx/0xCCBC00E33BFD9B9A766741BB9E26AC4C49FFF608E01291F19FBF821C18FF3D06}{0xCCBC\dots 3D06} & BTC.BTC & \href{https://btcscan.org/tx/2839C90D1CEEB5FFED317A845D6280312F6EB95E73C178677C6E1B790A89CFE5}{2839\dots CFE5} & 6.9313 \\
10 & ETH.ETH & \href{https://etherscan.io/tx/0xC8906F733F55A96C21D4E9B271EEC5AB92C04BE44409A93027C1A3D99151D2BE}{0xC890\dots D2BE} & BTC.BTC & \href{https://btcscan.org/tx/DCCA3D855FD34E2C35FA67C15C0087EBC9DD5B08EF682FB0ABAFA781ABC308BF}{DCCA\dots 08BF} & 6.9302 \\
\bottomrule
\end{tabular}
}
\end{table}

\section{THORChain cross-chain attack}
\label{app:thorchain-one-to-many}

\begin{figure}[t]
    \centering
    \includegraphics[width=1\linewidth]{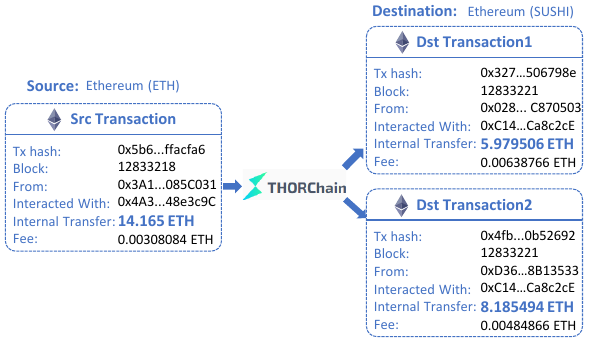}
    \caption{A representative one-to-two correspondence observed in the THORChain attack analysis.}
    \label{fig:thorchain-one-to-many}
\end{figure}

Figure~\ref{fig:thorchain-one-to-many} shows a representative one-to-two correspondence observed in the THORChain attack analysis. The source-side transaction was initiated on Ethereum and transferred 14.165 ETH into THORChain. Instead of producing a single destination-side settlement transaction, the flow was split into two destination-side transactions on Ethereum: one transferring 5.979506 ETH and the other transferring 8.185494 ETH. The two outgoing transfers occurred in the same destination block and interacted with the same destination-side contract. Their amounts sum to the source-side internal transfer amount, excluding transaction fees.

This case illustrates why explorer-level mappings may be incomplete. A public bridge interface may display only one outbound transaction or collapse the transfer into a simplified status view, thereby omitting part of the asset-distribution path. \textsc{XSplicer} does not rely on the explorer display logic. It instead reconstructs the correspondence from protocol-level hard evidence, including the THORChain \texttt{memo} and the Inbound/Outbound \texttt{Tx ID} linkage. 

\begin{figure}[t]
    \centering
    \includegraphics[width=0.95\linewidth]{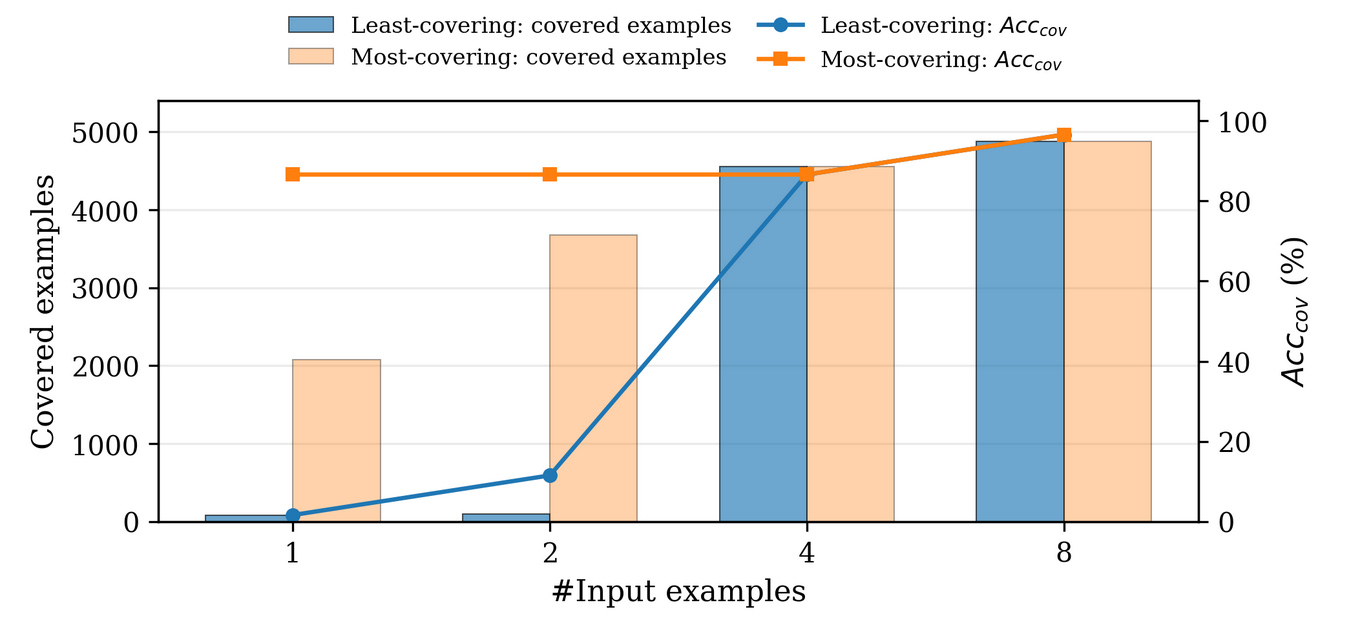}
    \caption{Ablation study on system inputs with varying numbers of cross-chain transaction pairs sampled via different strategies.}
    \label{fig:ablation}
\end{figure}

\begin{table}[t]
\centering
\caption{Protocol-specific evidence rules generated by \textsc{XSplicer} across the seven evaluated bridges.}
\label{tab:expert-validation-evidence}
\scriptsize
\setlength{\tabcolsep}{3pt}
\renewcommand{\arraystretch}{1}

\begin{tabularx}{\columnwidth}{
    @{}
    >{\raggedright\arraybackslash}p{0.16\columnwidth}
    >{\raggedright\arraybackslash}X
    @{}
}
\toprule
\textbf{Bridge} & \textbf{Hard Evidence Rules} \\
\midrule

\textbf{Allbridge}
&
\textbf{Destination:}
Chain from \texttt{destinationChainId}, \texttt{destinationDomain},
or \texttt{dst\_chain\_id}; recipient from
\texttt{mintRecipient}/\texttt{recipient}, or recipient bytes in
\texttt{SwapAndBridge}.
\textbf{Shared Evidence:}
Two-tiered strong evidence:
\texttt{message\_id}/\texttt{message\_hash}
$\rightarrow$ CCTP \texttt{nonce};
otherwise, recipient, token, amount, and time as soft evidence.
\\
\midrule

\textbf{deBridge}
&
\textbf{Destination:}
Chain from \texttt{CreatedOrder.order.takeChainId};
recipient candidates:
\texttt{receiverDst},
\texttt{orderAuthorityAddressDst}, and
\texttt{allowedTakerDst}
(often the target-side account for Solana).
\textbf{Shared Evidence:}
An exact \texttt{order\_id} match on both sides directly confirms the pair;
otherwise amount/time soft matching with capped confidence.
\\
\midrule

\textbf{Orbiter}
&
\textbf{Destination:}
Chain from \texttt{c}/\texttt{dstChainId} in
\texttt{memo}/\texttt{extData};
recipient from \texttt{t}, \texttt{to}, \texttt{recipient},
\texttt{receiver}, or \texttt{dstreceiver}.

\textbf{Shared Evidence:}
Source transaction identifier is strongest:
\texttt{h=<solana\_signature>} for Solana$\rightarrow$EVM;
for EVM$\rightarrow$Solana, check
\texttt{0x<evm\_txhash>} or \texttt{h=<evm\_txhash>}
in Solana \texttt{memo}/\texttt{obt}.
Otherwise recipient + amount/time.
\\
\midrule

\textbf{Symbiosis}
&
\textbf{Destination:}
Chain/recipient from \texttt{dstChain} and \texttt{dstAddress}
in \texttt{SwapToken}; currently \texttt{dstChain=5} denotes Solana.
Fallback fields include \texttt{dstAddress}, \texttt{recipient},
and \texttt{\_receiver}.

\textbf{Shared Evidence:}
No stable shared identifier; pairing uses the exact intersection of the
source-recovered Solana recipient and destination-side recipient,
with amount as auxiliary evidence.
\\
\midrule

\textbf{THORChain}
&
\textbf{Destination:}
Chain/recipient from THOR memo structures
(\texttt{SWAP/=:\!}, asset-tail, and channel-tail address clues).

\textbf{Shared Evidence:}
\texttt{OUT:<source\_tx\_hash>} is strongest;
\texttt{transferOut}/\texttt{router} semantics, memo structure,
and destination-address/amount relations provide strong auxiliary evidence.
\\
\midrule

\textbf{WanBridge}
&
\textbf{Destination:}
Chain from \texttt{tokenPairID} and, for CCTP,
\texttt{destinationDomain};
recipient from \texttt{userAccount}, \texttt{toAccount},
\texttt{receiver}, \texttt{toAddress}, or \texttt{mintRecipient}.
BTC$\rightarrow$EVM may additionally use candidate bridge contracts.

\textbf{Shared Evidence:}
\texttt{uniqueId} and source hash in Bitcoin \texttt{OP\_RETURN}
are strongest;
\texttt{tokenPairID}, \texttt{smgID}, and \texttt{userAccount}
are strong auxiliary evidence;
amount/time are weak evidence.
BTC$\rightarrow$EVM relies more on amount when structured identifiers
are absent.
\\
\midrule

\textbf{Wormhole}
&
\textbf{Destination:}
Chain from \texttt{recipientChain}/\texttt{toChain};
recipient from \texttt{recipient}/\texttt{recipient\_b32};
NTT directly uses
\texttt{recipientChain}+\texttt{recipient}.

\textbf{Shared Evidence:}
Core identifier:
\texttt{message\_id=(emitter\_chain\_id, emitter\_address, sequence)};
CCTP additionally uses
\texttt{messageBody}, \texttt{nonce}, and \texttt{sourceDomain}.
\\

\bottomrule
\end{tabularx}
\end{table}

\section{Impact of Public Examples}
\label{app:impact_of_examples}

We further evaluate how the number of public examples affects protocol knowledge reconstruction. Using Allbridge ETH$\rightarrow$SOL as a representative case, we compare different numbers of input examples over 4,999 test samples. We consider two extreme selection strategies: the least-covering strategy selects, at each step, the example that covers the fewest remaining similar-pattern samples, simulating an unfavorable selection; the most-covering strategy selects the example that covers the most similar-pattern samples, estimating the upper bound provided by representative examples.

\section{\rev{Protocol-Specific Evidence Rules Generated by \textsc{XSplicer}}}
\label{app:data:expert-validation}
\rev{TABLE~\ref{tab:expert-validation-evidence} summarizes the protocol-specific rules reconstructed in Stage I and compiled into the \texttt{Analyzer} and \texttt{Verifier} in Stage II for the seven-bridge benchmark. It lists the destination-side constraints used to bound candidate sets and the cross-chain evidence used to assess candidate correspondences.}

\section{Stage I Evaluation Metrics}
\label{appendix:metrics}

\rev{Cross-chain bridge protocols often use different terms for semantically equivalent entities, such as \texttt{recipient} and \texttt{to\_address}. We therefore define a semantic equivalence mapping operator $\sigma$ to align the predicted set $\mathcal{S}$ with the ground-truth specification set $\mathcal{G}$ (Gold Specs), using a predefined dictionary of protocol synonyms for reproducibility. We evaluate knowledge reconstruction using three metrics:}


\textbf{Destination-spec correctness:}
This metric evaluates the capability of the system to identify destination chain boundaries and address conventions. We compute the F1 score between the predicted field set $\mathcal{S}_{dst}$ and the baseline set $\mathcal{G}_{dst}$ as follows:
\begin{equation}
    \text{F1}_{dst} = \frac{2 \cdot |\mathcal{S}_{dst} \cap_{\sigma} \mathcal{G}_{dst}|}{|\mathcal{S}_{dst}| + |\mathcal{G}_{dst}|}
\end{equation}
where $\cap_{\sigma}$ denotes the semantic intersection under the mapping operator $\sigma$. This metric facilitates the assessment of the system's confidence in delineating on-chain search boundaries.

\textbf{Cross-chain shared-spec correctness:}
\rev{This metric measures the recovery of protocol invariants and relational mappings, such as cross-chain transaction identifiers and value-conversion formulas:}
\begin{equation}
    \text{F1}_{cross} = \frac{2 \cdot |\mathcal{S}_{cross} \cap_{\sigma} \mathcal{G}_{cross}|}{|\mathcal{S}_{cross}| + |\mathcal{G}_{cross}|}
\end{equation}
\rev{The overlap between $\mathcal{S}_{cross}$ and $\mathcal{G}_{cross}$ reflects how accurately Stage I recovers the invariants and mappings used by downstream verifiers.}

\textbf{Groundedness (Gnd):}
\rev{Groundedness measures whether semantically correct predictions are supported by the provided evidence rather than model priors. Let $\mathcal{H} = \mathcal{S} \cap_{\sigma} \mathcal{G}$ denote the set of correct predictions. We divide $\mathcal{H}$ by evidence provenance into:}
\textcircled{\small 1} $\mathcal{H}_{strong}$: \rev{directly supported by explicit text or transaction JSON (Direct Evidence);}
\textcircled{\small 2} $\mathcal{H}_{weak}$: \rev{supported by partial contextual clues requiring moderate inference (Partial Evidence);}
\textcircled{\small 3} $\mathcal{H}_{none}$: \rev{unsupported by the provided context and thus attributable to model-internal knowledge (Hallucination).}
\rev{The \textbf{Gnd} metric is the proportion of correct predictions supported by verifiable evidence:}
\begin{equation}
    \text{Gnd} = \frac{|\mathcal{H}_{strong} \cup \mathcal{H}_{weak}|}{|\mathcal{H}|}.
\end{equation}

\newcommand{\cmark}{\ding{51}}
\newcommand{\xmark}{\ding{55}}

\begin{table*}[t]
\centering
\caption{Surveyed cross-chain bridges, documentation, explorers, and crawlability.}
\label{tab:bridge-survey-two-panel}
\scriptsize
\setlength{\tabcolsep}{1.4pt}
\renewcommand{\arraystretch}{0.95}

\begin{minipage}[t]{0.495\textwidth}
\centering
\begin{tabular}{@{}r >{\raggedright\arraybackslash}p{0.29\linewidth} >{\raggedright\arraybackslash}p{0.30\linewidth} c c >{\raggedright\arraybackslash}p{0.12\linewidth}@{}}
\toprule
\# & \makecell[l]{Bridge /\\Docs} & Explorer & \makecell{EVM/BTC/\\SOL} & API & Note \\
\midrule
1 & \href{https://docs.across.to/}{Across} & -- & -- & -- & -- \\
2 & \href{https://everscale.network/blog/dexada/}{Adaever} & -- & -- & -- & -- \\
\rowcolor{gray!10} 
3 & \href{https://docs-core.allbridge.io/}{Allbridge Core} & \href{https://core.allbridge.io/explorer}{Allbridge Explorer} & \cmark & \cmark & \textbf{Allbridge} \\
4 & \href{https://www.altitudedefi.com/}{Altitude} & -- & -- & -- & -- \\
5 & \href{https://stargateprotocol.gitbook.io/stargate/user-docs/the-aptosbridge/the-aptosbridge}{Aptos Bridge} & -- & -- & -- & -- \\
6 & \href{https://docs.arbitrum.io/for-users/bridging/bridging-overview}{Arbitrum Bridge} & \href{https://bridge.arbitrum.io/}{Arbitrum Bridge Explorer} & \xmark & -- & -- \\
7 & \href{https://docs.base.org/base-chain/network-information/bridges-mainnet}{Base Bridge} & -- & -- & -- & -- \\
8 & \href{https://www.binance.com/en/blog/ecosystem/421499824684903626}{Binance Bridge 2.0} & -- & -- & -- & -- \\
9 & \href{https://docs.boba.network/basics/bridging/index}{Boba Gateway} & -- & -- & -- & -- \\
10 & \href{https://github.com/wormhole-foundation}{BoringDAO oPortal} & -- & -- & -- & -- \\
11 & \href{https://github.com/celer-network}{Celer cBridge} & \href{https://celerscan.com/}{Celer cBridge Explorer} & \xmark & -- & -- \\
12 & \href{https://docs.chainport.io/}{ChainPort} & \href{https://app.chainport.io/explore}{ChainPort Explorer} & \xmark & -- & -- \\
13 & \href{https://docs.chainspot.io/}{Chainspot} & -- & -- & -- & -- \\
14 & \href{https://docs.connext.network/}{Connext Bridge} & -- & -- & -- & -- \\
15 & \href{https://core.app/discover/education/avalanche-bridge}{Core Avalanche Bridge} & -- & -- & -- & -- \\
16 & \href{https://docs.coredao.org/docs/Dev-Guide/core-bridge-resources}{Core Bitcoin Bridge} & -- & -- & -- & -- \\
17 & \href{https://counterstake.org/how-it-works}{Counterstake Bridge} & -- & -- & -- & -- \\
18 & \href{https://docs.bnbchain.org/bnb-smart-chain/cross-chain-bridge/}{Cross-Chain Bridge} & -- & -- & -- & -- \\
19 & \href{https://docs.croxswap.com/guide/bsc-heco-bridge}{CroxSwap Bridge} & -- & -- & -- & -- \\
\rowcolor{gray!10} 
20 & \href{https://github.com/debridge-finance}{DAFI dBridge} & \href{https://app.debridge.com/orders}{DAFI dBridge Explorer} & \cmark & \cmark & \textbf{deBridge} \\
21 & \href{https://decimalchain.com/blog/ibc-and-cross-chain-bridges/}{Decimalchain} & \href{https://explorer.decimalchain.com/}{Decimalchain Explorer} & \xmark & -- & -- \\
22 & \href{https://debridge.com/learn/blog/deswap-widget-a-turnkey-cross-chain-solution-for-web3/}{deSwap} & -- & -- & -- & -- \\
\rowcolor{gray!10} 
23 & \href{https://github.com/debridge-finance}{DLN (Powered by deBridge)} & \href{https://app.debridge.com/orders}{DLN (Powered by deBridge) Explorer} & \cmark & \cmark & \textbf{deBridge} \\
24 & \href{https://docs.elk.finance/}{ElkNet Bridge} & -- & -- & -- & -- \\
25 & \href{https://docs.evodefi.com/}{EVODeFi Token Bridge} & -- & -- & -- & -- \\
26 & \href{https://docs.eywa.fi/eywa-ecosystem/products/eywa-v1/cross-chain-liquidity-protocol/eywa-token-bridge}{EYWA Token Bridge} & -- & -- & -- & -- \\
27 & \href{https://docs.floki.com/whitepaper/tokenomics/multi-chain-protocol}{Floki Bridge} & -- & -- & -- & -- \\
28 & \href{https://docs.glitterfinance.org/}{Glitter Bridge} & -- & -- & -- & -- \\
29 & \href{https://github.com/Gravity-Bridge}{Gravity Bridge} & -- & -- & -- & -- \\
30 & \href{https://github.com/harmony-one/horizon}{Harmony Bridge} & -- & -- & -- & -- \\
31 & \href{https://hashbon.gitbook.io/hashbon-rocket/}{Hashbon Rocket} & -- & -- & -- & -- \\
32 & \href{https://docs.helixbox.ai/labs/}{Helix Bridge} & -- & -- & -- & -- \\
33 & \href{https://docs.holograph.xyz/}{Holograph} & -- & -- & -- & -- \\
34 & \href{https://docs.hop.exchange/}{Hop.Exchange} & -- & -- & -- & -- \\
35 & \href{https://docs.hotcross.com/}{Hot Cross Multi-Chain Bridge} & -- & -- & -- & -- \\
36 & \href{https://github.com/bcnmy}{Hyphen} & -- & -- & -- & -- \\
37 & \href{https://docs.hyperbridge.network}{HyperBridge} & \href{https://explorer.hyperbridge.network/}{HyperBridge Explorer} & \xmark & -- & -- \\
38 & \href{https://docs.crosschainbridge.org/}{iCrosschain} & -- & -- & -- & -- \\
39 & \href{https://docs.tac.build/bridge}{Instant Cross} & -- & -- & -- & -- \\
40 & \href{https://docs.interlay.io/}{Interlay} & -- & -- & -- & -- \\
41 & \href{https://docs.interport.fi/}{Interport Finance} & \href{https://explorer.interport.fi/}{Interport Finance Explorer} & \xmark & -- & -- \\
42 & \href{https://docs.iotex.io/blockchain/ecosystem/iotube-bridge}{ioTube} & -- & -- & -- & -- \\
43 & \href{https://docs.li.fi/}{Jumper Exchange} & -- & -- & -- & -- \\
44 & \href{https://docs.kcc.io/individuals/bridge-assets/kcc-bridge}{KCC Bridge} & -- & -- & -- & -- \\
45 & \href{https://kccpad.medium.com/how-to-set-up-your-metamask-for-the-kcc-network-1d4d14b4d560}{KCCPad Bridge} & -- & -- & -- & -- \\
46 & \href{https://github.com/interlay/interbtc-ui}{Kintsugi} & -- & -- & -- & -- \\
47 & \href{https://docs.layerswap.io/}{Layerswap} & -- & -- & -- & -- \\
48 & \href{https://docs.magpiexyz.io/bridge}{Magpie Protocol} & -- & -- & -- & -- \\
49 & \href{https://docs.manta.network/docs/manta-pacific/Tools/Bridge/Native\%20Bridge/How\%20to\%20Use\%20Native\%20Bridge}{Manta Pacific Bridge} & -- & -- & -- & -- \\
50 & \href{https://docs.mantle.xyz/network/user-guides/bridge}{Mantle Bridge} & -- & -- & -- & -- \\
51 & \href{https://docs.mayan.finance/}{Mayan Finance} & \href{https://swap.mayan.finance/}{Mayan Finance Explorer} & \cmark & \xmark & -- \\
52 & \href{https://docs.meson.fi/}{Meson} & \href{https://explorer.meson.fi/}{Meson Explorer} & \xmark & -- & -- \\
53 & \href{https://docs.meter.io/learn/passport}{Meter Passport} & -- & -- & -- & -- \\
54 & \href{https://www.metis.io/bridge}{Metis Bridge} & -- & -- & -- & -- \\
55 & \href{https://docs.curvegrid.com/multibaas/}{MultiBaas Bridge BETA} & \href{https://app.lighter.xyz/explorer}{MultiBaas Explorer} & \xmark & -- & -- \\
56 & \href{https://github.com/anyswap}{Multichain (previously Anyswap)} & \href{https://scan.multichain.org/\#/}{Multichain Explorer} & \cmark & \xmark &  \\
57 & \href{https://github.com/MixinNetwork/bridge.mvm.app}{MVM Bridge} & -- & -- & -- & -- \\
58 & \href{https://github.com/celer-network}{NFT OmniBridge} & -- & -- & -- & -- \\
59 & \href{https://github.com/nomad-xyz}{Nomad} & -- & -- & -- & -- \\
60 & \href{https://github.com/broxus}{Octus Bridge} & -- & -- & -- & -- \\
61 & \href{https://www.okx.com/help/oktc-will-discontinue-the-ibc-transfer-product}{OKX IBC Transfer} & -- & -- & -- & -- \\
62 & \href{https://docs.gnosischain.com/bridges/about-token-bridges/omnibridge}{OmniBridge} & -- & -- & -- & -- \\
63 & \href{https://docs.omnisea.org/omnichain-router/overview}{Omnisea} & -- & -- & -- & -- \\
64 & \href{https://docs.bnbchain.org/bnb-opbnb/developers/bep20-crosschain/}{opBNB Bridge} & -- & -- & -- & -- \\
65 & \href{https://doc.openswap.xyz/}{OpenSwap} & -- & -- & -- & -- \\
66 & \href{https://docs.optimism.io/app-developers/guides/bridging/standard-bridge}{Optimism Bridge} & -- & -- & -- & -- \\
\bottomrule
\end{tabular}
\end{minipage}\hfill
\begin{minipage}[t]{0.495\textwidth}
\centering
\begin{tabular}{@{}r >{\raggedright\arraybackslash}p{0.29\linewidth} >{\raggedright\arraybackslash}p{0.30\linewidth} c c >{\raggedright\arraybackslash}p{0.12\linewidth}@{}}
\toprule
\# & \makecell[l]{Bridge /\\Docs} & Explorer & \makecell{EVM/BTC/\\SOL} & API & Note \\
\midrule
67 & \href{https://github.com/orbit-chain}{Orbit Bridge} & -- & -- & -- & -- \\
\rowcolor{gray!10} 
68 & \href{https://github.com/Orbiter-Finance}{Orbiter Finance} & \href{https://www.orbiter.finance/explore}{Orbiter Finance Explorer} & \cmark & \cmark & \textbf{Orbiter} \\
69 & \href{https://docs.orion.xyz/}{Orion Bridge} & -- & -- & -- & -- \\
70 & \href{https://docs.owlto.finance/}{Owlto} & -- & -- & -- & -- \\
71 & \href{https://doc.zyx.network/zyx-docs/bridge-pandorum.io}{Pandorum BETA} & -- & -- & -- & -- \\
72 & \href{https://github.com/crypto-pepe/bridge-evm-contracts}{PepeBridge} & \href{https://bridge.pepe.team/explorer/}{PepeBridge Explorer} & \xmark & -- & -- \\
73 & \href{https://github.com/pnetwork-association}{Pheasant Network} & -- & -- & -- & -- \\
74 & \href{https://github.com/pnetwork-association}{pNetwork dApp} & -- & -- & -- & -- \\
75 & \href{https://polynetwork.medium.com/notice-of-complete-termination-of-poly-network-services-3470ef78d9d9}{PolyBridge} & -- & -- & -- & -- \\
76 & \href{https://docs.polygon.technology/tools/wallets/portal/}{Polygon Portal} & -- & -- & -- & -- \\
77 & \href{https://docs.polygon.technology/zkEVM/}{Polygon zkEVM Bridge} & -- & -- & -- & -- \\
\rowcolor{gray!10} 
78 & \href{https://docs.wormhole.com/wormhole/}{Portal Token Bridge (previously Wormhole)} & \href{https://wormholescan.io/}{Wormhole Explorer} & \cmark & \cmark & \textbf{Wormhole} \\
79 & \href{https://prom.gitbook.io/prom}{PROM Bridge} & -- & -- & -- & -- \\
80 & \href{https://docs.radar.global/}{Radar Bridge} & \href{https://app.hyperliquid.xyz/explorer}{Radar Bridge Explorer} & \xmark & -- & -- \\
81 & \href{https://doc.aurora.dev/bridge/introduction/}{Rainbow Bridge} & -- & -- & -- & -- \\
82 & \href{https://docs.relay.link/}{Relay Bridge} & -- & -- & -- & -- \\
83 & \href{https://docs.relay.link/what-is-relay}{Relay Chain Bridge} & -- & -- & -- & -- \\
84 & \href{https://bridge.renproject.io/}{RenBridge} & -- & -- & -- & -- \\
85 & \href{https://api-docs.retrobridge.io/}{Retrobridge} & -- & -- & -- & -- \\
86 & \href{https://docs.rhino.fi/}{rhino.fi} & -- & -- & -- & -- \\
87 & \href{https://docs.roninchain.com/apps/bridge}{Ronin Bridge} & -- & -- & -- & -- \\
88 & \href{https://docs.routerprotocol.com/develop/asset-transfer-via-nitro/high-level-workflow/}{Router Nitro} & -- & -- & -- & -- \\
89 & \href{https://dev.rootstock.io/rsk/rbtc/concepts/token-bridge/}{RSK Token Bridge} & -- & -- & -- & -- \\
90 & \href{https://docs.rubic.finance/}{Rubic} & -- & -- & -- & -- \\
91 & \href{https://safecoin.org/safebridge}{SafeBridge} & -- & -- & -- & -- \\
92 & \href{https://docs.axelar.dev/resources/satellite}{Satellite by Axelar} & -- & -- & -- & -- \\
93 & \href{https://docs.scroll.io/en/user-guide/bridging/bridge/}{Scroll Bridge} & -- & -- & -- & -- \\
94 & \href{https://medium.com/secret-network-ecosystem-and-technology/how-to-use-the-secret-tunnel-a36cebfd4949}{Secret Tunnel} & -- & -- & -- & -- \\
95 & \href{https://docs.sifchain.network/}{Sifchain Bridge} & -- & -- & -- & -- \\
96 & \href{https://github.com/wormhole-foundation}{SKALE Portal Bridge} & \href{https://wormholescan.io/}{SKALE Portal Bridge Explorer} & \xmark & -- & -- \\
97 & \href{https://docs.solarbeam.io/getting-started/bridge}{SolarBeam Bridge} & -- & -- & -- & -- \\
98 & \href{https://wiki.sovryn.app/en/sovryn-dapp/bridge}{Sovryn Bridge} & -- & -- & -- & -- \\
99 & \href{https://soy-finance.gitbook.io/soy-finance/}{SOY Bridge} & -- & -- & -- & -- \\
100 & \href{https://docs.spooky.fi/getting-started/bridge-fantom}{SpookySwap Bridge} & -- & -- & -- & -- \\
101 & \href{https://docs.squidrouter.com/}{Squid} & -- & -- & -- & -- \\
102 & \href{https://docs.stargate.finance/}{Stargate} & -- & -- & -- & -- \\
103 & \href{https://docs.sushi.com/}{SushiSwap} & -- & -- & -- & -- \\
104 & \href{https://docs.suterusu.io/}{Suter Bridge ALPHA} & -- & -- & -- & -- \\
105 & \href{https://github.com/swim-io/swim/tree/master/docs}{Swim Protocol} & -- & -- & -- & -- \\
\rowcolor{gray!10} 
106 & \href{https://docs.symbiosis.finance/}{Symbiosis} & \href{https://symbiosis.finance/}{Symbiosis Explorer} & \cmark & \cmark & \textbf{Symbiosis} \\
107 & \href{https://docs.synapseprotocol.com/docs/Bridge/}{Synapse Protocol} & -- & -- & -- & -- \\
108 & \href{https://x.com/terra\_money/status/1853823983369818246}{Terra Bridge} & -- & -- & -- & -- \\
109 & \href{https://github.com/consenlabs}{The Voyager} & -- & -- & -- & -- \\
\rowcolor{gray!10} 
110 & \href{https://docs.thorswap.finance/}{THORSwap} & \href{https://thorchain.net/dashboard}{THORSwap Explorer} & \cmark & \cmark & \textbf{THORChain} \\
111 & \href{https://docs.developers.thundercore.com/product-protocol/bridges}{ThunderCore Bridge} & -- & -- & -- & -- \\
112 & \href{https://docs.timewarp.finance/how-to-guides/bridge-how-to-swap-time-token-between-ethereum-bsc-and-polygon}{TIME Bridge} & -- & -- & -- & -- \\
113 & \href{https://blog.ton.org/ton-retires-toncoin-bridge}{TON Bridge} & -- & -- & -- & -- \\
114 & \href{https://docs.transit.finance/}{Transit Swap} & \href{https://explorer.transit.finance/?type=transit\#/}{Transit Swap Explorer} & \cmark & \xmark & -- \\
115 & \href{https://medium.com/@trantornetwork/trantor-litepaper-24de2b763980}{Trantor} & -- & -- & -- & -- \\
116 & \href{https://tronpad.network/}{TronPad} & -- & -- & -- & -- \\
117 & \href{https://docs.txsync.io/product-integration/bridge}{txSync Bridge} & -- & -- & -- & -- \\
118 & \href{https://github.com/umbrella-network/technical-documentation}{Umbrella Token Bridge} & -- & -- & -- & -- \\
119 & \href{https://legacy.umbria.network/docs/docs-page}{Umbria Narni Bridge} & -- & -- & -- & -- \\
120 & \href{https://github.com/viaprotocol/via-protocol-info}{Via Protocol Exchange} & -- & -- & -- & -- \\
121 & \href{https://docs.vires.finance/}{Vires.Finance} & -- & -- & -- & -- \\
122 & \href{https://docs.voltage.finance/voltage/the-platform/bridge}{Voltage Bridge} & -- & -- & -- & -- \\
123 & \href{https://docs.voltswap.finance/how-to-bridge-funds}{VoltSwap} & -- & -- & -- & -- \\
\rowcolor{gray!10} 
124 & \href{https://docs.wanchain.org/products/wanbridge}{WanBridge} & \href{https://www.wanscan.org/}{WanBridge Explorer} & \cmark & \cmark & \textbf{WanBridge} \\
125 & \href{https://docs.wax.io/learn/about-wax/wax-interoperability}{WAX ETH Bridge} & -- & -- & -- & -- \\
126 & \href{https://docs.gnosischain.com/bridges/About\%20Token\%20Bridges/xdai-bridge}{xDai Bridge} & -- & -- & -- & -- \\
127 & \href{https://docs.xp.network/}{XP.NETWORK NFT Multi-Chain Bridge} & -- & -- & -- & -- \\
128 & \href{https://github.com/Glitter-Finance}{XY Finance} & -- & -- & -- & -- \\
129 & \href{https://help.zapper.xyz/hc/en-us/sections/7815037518865-Bridge}{Zapper} & -- & -- & -- & -- \\
130 & \href{https://docs.zigzag.exchange/}{ZigZag Bridge} & -- & -- & -- & -- \\
131 & \href{https://github.com/celer-network}{zkBridge} & -- & -- & -- & -- \\
  &   &  &  &  &   \\
\bottomrule
\end{tabular}
\end{minipage}

\vspace{0.25em}
\begin{minipage}{0.98\textwidth}
\footnotesize
\emph{Note.} The bridge name is hyperlinked to its documentation. When an explorer is available, the explorer column is hyperlinked and displayed as ``Bridge Explorer''. \cmark{} and \xmark{} denote available and unavailable support, respectively; ``--'' denotes unreported information.
\end{minipage}
\end{table*}

\end{document}